\documentclass[twoside,twocolumn,a4paper]{IEEEtran}
\usepackage{url}
\usepackage{mathtools}
\usepackage{amsfonts}
\usepackage{setspace}
\usepackage{mathptmx}       
\usepackage{helvet}         
\usepackage{courier}       
\usepackage{type1cm}       
\usepackage{makeidx}
\usepackage[bottom]{footmisc}
\usepackage{multicol}
\usepackage{amsmath}
\makeindex

\def\mat#1{{\bf #1}}

\newcommand{\be}{\begin{equation}}
\newcommand{\ee}{\end{equation}}
\newcommand{\rn}{{\rm n}}
\newcommand{\re}{{\rm e}}

\begin{document}

\title{The Potential Energy Surface in molecular quantum mechanics}
\author {Brian Sutcliffe\thanks{B.T. Sutcliffe, Service de Chimie quantique et Photophysique, Universit\'{e} Libre de Bruxelles, B-1050 Bruxelles, Belgium. {email: bsutclif@ulb.ac.be}},~~R. Guy Woolley\thanks{R.G. Woolley, School of Science and Technology, Nottingham Trent University, Nottingham NG11 8NS, U.K.}
\thanks{Manuscript accepted by \textit{Progress in Theoretical Chemistry and Physics, March 2013.}}}

\maketitle

\begin{abstract} The idea of a \index{Potential Energy Surface}Potential Energy Surface (PES) forms the basis of almost all accounts of the mechanisms of chemical reactions, and much of theoretical molecular spectroscopy. It is assumed that, in principle, the PES can be calculated by means of \index{clamped-nuclei} clamped-nuclei \index{electronic structure}electronic structure calculations based upon the \index{Schr\"{o}dinger}Schr\"{o}dinger \index{Coulomb}Coulomb \index{Hamiltonian}Hamiltonian. This article is devoted to a discussion of the origin of the idea, its development in the context of the \index{Old Quantum Theory}Old Quantum Theory, and its present status in the quantum mechanics of molecules. It is argued that its present status must be regarded as uncertain.
\end{abstract}
\pagestyle{plain}
\section{Introduction}
\label{Intro}
\begin{quote}
The \index{Coulomb}Coulombic \index{Hamiltonian}Hamiltonian $\mathsf{H}'$ does not provide much obvious information or guidance, since there is [\textit{sic}] no specific assignments of the electrons occurring in the systems to the atomic nuclei involved - hence there are no atoms, isomers, conformations etc. In particular one sees no \textit{molecular symmetry}, and one may even wonder where it comes from. Still it is evident that all of this information must be contained somehow in the \index{Coulomb}Coulombic \index{Hamiltonian}Hamiltonian $\mathsf{H}'$ \cite{POL:89}.
\end{quote}
Per-Olov \index{L\"{o}wdin}L\"{o}wdin, Pure. Appl. Chem. {\bf 61}, p.2071, (1989)\\

This paper addresses the question \index{L\"{o}wdin}L\"{o}wdin wondered about in terms of what quantum mechanics has to says about molecules. A conventional chemical description of a stable molecule is a collection of atoms held in a semi-rigid arrangement by chemical bonds, which is summarized as a \index{molecular structure}molecular structure. Whatever `chemical bonds' might be physically, it is natural to interpret this statement in terms of bonding forces which are conservative. Hence a stable molecule can be associated with a \textit{potential energy} function that has a minimum value below the energy of all the clusters that the molecule can be decomposed into. Finding out about these forces, or equivalently the associated potential energy, has been a major activity for the past century. There is no \textit{a priori} specification of \textit{atomic} interactions from basic physical laws so the approach has been necessarily indirect.

After the discovery of the electron \cite{JJT:99} and the triumph of the atomic, mechanistic view of the constitution of matter, it became universally accepted that any specific molecule consists of a certain number of electrons and nuclei in accordance with its chemical formula. This can be translated into a microscopic model of point charged particles interacting through \index{Coulomb}Coulomb's law with non-relativistic kinematics.  These assumptions fix the molecular \index{Hamiltonian}Hamiltonian as precisely what \index{L\"{o}wdin} L\"{o}wdin referred to as the `\index{Coulomb}Coulombic \index{Hamiltonian}Hamiltonian',
\begin{equation}
\mathsf{H}~=~\sum_i^n\frac{p_i^{2}}{2m_i}~+~\sum_{i < j}^{n}\frac{e_i e_j}{4
\pi \epsilon_0 |{\bf q}_i~-~{\bf q}_j|}
\label{molHam}
\end{equation}
where the $n$ particles are described by empirical charge and mass parameters \{$e_i,m_i~1~=~1,\ldots n$\}, and \index{Hamiltonian}Hamiltonian canonical variables \{${\bf q}_i,{\bf p}_i,~i~=~1\ldots n$\}, which after quantization are regarded as non-commuting operators.

As is well-known \index{classical dynamics}classical dynamics based on equation (\ref{molHam}) fails completely to account for the stability of atoms and molecules, as evidenced through the facts of chemistry and spectroscopy. And so, starting about a century ago, there was a progressive modification of dynamics as applied to the microscopic world from classical (`rational') mechanics, through the years of the \index{Old Quantum Theory}Old Quantum Theory until finally quantum mechanics was defined. This slow evolution left its mark on the development of molecular theory in as much that classical ideas survive in modern Quantum Chemistry. In the following sections we review some aspects of this progression; we also emphasize that a direct approach to a quantum theory of a molecule can be based on the quantized version of (\ref{molHam}), simply as an extension of the highly successful quantum theory of the atom.  

It is of interest to compare this so-called `\index{Isolated Molecule}Isolated Molecule' model with the conventional account; after all, the sentiment of the quotation from \index{L\"{o}wdin}L\"{o}wdin reflects the widespread view that the model is the fundamental basis of Quantum Chemistry. Even though there are no closed solutions for molecules, it is certainly possible to characterize important qualitative features of the solutions for the model because they are determined by the \textit{form} of the defining equations \cite{POL:89}, \cite{WS:03}, \cite{SW:12}. One of the most important ideas in molecular theory is the \index{Potential Energy Surface}Potential Energy Surface for a molecule; this is basic for theories of chemical reaction rates and for molecular spectroscopy. In \S \ref{class} we discuss some aspects of its classical origins. Then in \S \ref{quant} we revisit the same topics from the standpoint of quantum mechanics, where we will see that if we eschew the conventional classical input (classical fixed nuclei), \textit{there are no \index{Potential Energy Surface}Potential Energy Surfaces} in the solutions derived from (\ref{molHam}). It is \textit{not} the case that the conventional approach via the \index{clamped-nuclei} clamped-nuclei \index{Hamiltonian}Hamiltonian is merely a convenience that permits practical calculation (in modern terms, computation) with results concordant with the underlying \index{Isolated Molecule}Isolated Molecule model that would be obtained if only the computations could be done. On the contrary, a qualitative modification of the formalism is imposed by hand. The paper concludes in \S \ref{Disc} with a discussion of these results; some relevant mathematical results are illustrated in the Appendix. 

We wish to emphasize that the paper is about a difficult technical problem; it is not a contribution to the philosophy of science. In the traditional picture, equation (\ref{coupl}) is widely held to be exact in principle, so if the \index{adiabatic}adiabatic approximation is found to be inadequate we would expect to do `better' by including coupling terms. Our analysis implies that belief is not well founded because (\ref{coupl}) is not well founded \emph{a priori} in quantum mechanics; it requires an extra ingredient put in by hand.  It might work, or it might not; in other words it is not a sure-fire route to a better account. While we can't offer a better alternative, that information is surely important for chemical physics.

\section{Classical origins}
\label{class}
The idea of a \index{Potential Energy Surface}Potential Energy Surface can be glimpsed in the beginnings of chemical reaction rate theory that go beyond the purely thermodynamic considerations of van 't Hoff and \index{Duhem}Duhem more than a century ago, and in the first attempts to understand molecular (`band') spectra in dynamical terms in the same period. Thereafter progress was rapid as the newly emerging ideas of a `quantum theory' were developed; by the time that quantum mechanics was finalized (1925/6) ideas about the separability of electronic and nuclear motions in molecules were common currency, and were carried forward into the new era. In this section we describe how this development took place.

\subsection{Rates of chemical reactions - Ren\'{e} \index{Marcelin}Marcelin}
\label{chreact}
The idea of basing a theory of chemical reactions (\index{chemical dynamics}chemical dynamics) on an energy function that varies with the configurations of the participating molecules seems to be due to \index{Marcelin}Marcelin. In his last published work, his thesis, \cite{RM1}, \index{Marcelin}Marcelin showed how the Boltzmann distribution for a system in thermal equilibrium and statistical mechanics can be used to describe the rate, $v$, of a chemical reaction. The same work was republished in the Annales de Physique \cite{RM2} shortly after his 
death\footnote{Ren\'{e} \index{Marcelin}Marcelin was killed in action fighting for France in September 1914.} . The main conclusions of the thesis were summarized in two short notes published in Comptes Rendus in early 1914 \cite{RM3},\cite{RM4}. His fundamental result can be expressed, in modern terms, as
\begin{equation}
v~=~M\big(e^{-\Delta G_{+}^{\#}/RT}~-~e^{-\Delta G_{-}^{\#}/RT}\big)
\label{Meqn}
\end{equation}
where $R$ is the molar gas constant, $T$ is the temperature in Kelvin, the subscripts $+,-$ refer to the forward and reverse reactions, and $\Delta G^{\#}$ is the change in the molar \index{Gibbs}Gibbs (free) energy in going from the initial (+) or final (-) state to the `activated state'. The pre-exponential factor M is obtained formally from statistical mechanics. \index{Marcelin}Marcelin gave several derivations of this result using both thermodynamic arguments and also the statistical mechanics he had learnt from \index{Gibbs}Gibbs's famous memoir \cite{JWG}. It is perhaps worth remarking that \index{Gibbs}Gibbs saw statistical mechanics as the completion of \index{Newton}Newtonian mechanics through its extension to conservative systems with an arbitrarily large, though finite, number of degrees of freedom. The laws of thermodynamics could easily be obtained from the principles of statistical mechanics, of which they were the incomplete expression, but \index{Gibbs}Gibbs did not require thermodynamic systems to be made up of molecules; he explicitly did not wish his account of rational mechanics to be based on hypotheses concerning the constitution of matter, which at the time were still controversial \cite{LNa}.

>From our point of view the most interesting aspect of \index{Marcelin}Marcelin's account is the suggestion that molecules can have more degrees of freedom than those of simple point material particles. In this perspective, a molecule can be assigned a set of Lagrangian coordinates ${\bf q}~=~q_1,q_2, \ldots, q_n,$ and their corresponding canonical momenta ${\bf p}~=~ p_1,p_2, \ldots p_n$. Then the instantaneous \textit{state} of the molecule is associated with a `representative' point in the canonical \index{phase-space}phase-space ${\cal{P}}$ of dimension $2n$, and so ``as the position, speed or structure of the molecule changes, its representative point traces a trajectory in the $2n$-dimensional \index{phase-space}phase-space'' \cite{RM1}.

In his \index{phase-space}phase-space representation of a chemical reaction the transformation of reactant molecules into product molecules was viewed in terms of the passage of a set of trajectories associated with the `active' molecules through a `critical surface' ${\cal{S}}$ in ${\cal{P}}$ that divides ${\cal{P}}$ into two parts, one part being associated with the reactants, the other with the products. Such a [hyper]surface is defined by a relation 
\begin{displaymath}
S({\bf q},{\bf p})~=~0 
\end{displaymath}
According to \index{Marcelin}Marcelin, for passage through this surface it is required\footnote{that a molecule must reach a certain region of space at a suitable angle, that its speed must exceed a certain limit, that its internal structure must correspond to an unstable configuration \textit{etc};~....} \cite{RM1}
\begin{quote}
[une mol\'{e}cule] il faudra [....] qu'elle atteigne une certaine r\'{e}gion de l'\'{e}space sous une obliquit\'{e} convenable, que sa vitesse d\'{e}passe une certain limite, que sa structure interne corresponde \`{a} une configuration instable, \textit{etc.}; ~....
\end{quote}

In modern notation, the volume of a cell in the $2n$-dimensional \index{phase-space}phase-space is
\begin{displaymath}
d\varpi ~=~d{\bf q}~d{\bf p}
\end{displaymath}
The number of points in $d\varpi$ is given by the \index{Gibbs}Gibbs distribution  function $f$
\begin{equation}
d\nu~=~f({\bf q},{\bf p},t)~d\varpi
\label{equ}
\end{equation}
\index{Marcelin}Marcelin chose the distribution function for the active molecules as
\begin{equation}
f({\bf q},{\bf p},t)~=~e^{-G_{+}^{\#}/RT}~e^{-{\cal{H}}({\bf q},{\bf p})/k_BT}
\label{dist}
\end{equation}
where $k_B$ is Boltzmann's constant, ${\cal{H}}$ is the \index{Hamiltonian}Hamiltonian function for the molecule, and $G_{+}^{\#}$ is the Gibb's free energy of the active molecules relative to the mean energy of the reactant molecules. It is independent of the canonical variables. There is an analogous expression for the reverse reaction involving $G_{-}^{\#}$. \index{Marcelin}Marcelin quoted a formula due to \index{Gibbs}Gibbs \cite{JWG} for the number of molecules $dN$ crossing a surface element $ds$ in the critical surface in the neighbourhood of ${\bf q},{\bf p}$, in time $dt$, which may be written in shorthand as
\begin{displaymath}
dN~=~dt~f({\bf q},{\bf p},t)~J(\dot{\bf q},\dot{\bf p},{\bf q},{\bf p}) 
\end{displaymath}
where $\dot{\bf p}, \dot{\bf q}$ are regarded as functions of ${\bf q}, {\bf p}$ by virtue of Hamilton's equations of motion. The total rate is
\begin{equation}
v~=~\int d\varpi~f({\bf q},{\bf p})~J~\delta[S({\bf q},{\bf p})]
\label{flux}
\end{equation}
where the delta function confines the integration to the critical surface ${\cal{S}}$. Equation (\ref{Meqn}) results from taking the difference between this expression for the forward and reverse reactions, and factoring out the terms in $G_{\pm}^{\#}$; the remaining integration, which \index{Marcelin}Marcelin did not evaluate, defines the multiplying factor $M$.

\subsection{Molecular spectroscopy and the \index{Old Quantum Theory}Old Quantum Theory}
\label{OQT}
Although the discussion in the previous section looks familiar, it does so only because of the modern interpretation we put upon it\footnote{Nevertheless it seems proper to regard \index{Marcelin}Marcelin's introduction of \index{phase-space}phase-space variables and a critical reaction surface into \index{chemical dynamics}chemical dynamics as the beginning of a formulation of the Transition State Theory that was developed by Wigner in the 1930's \cite{EPW1} - \cite{WSW}. The $2n$ \index{phase-space}phase-space variables ${\bf q},{\bf p}$ were identified with the $n$ \textit{nuclei} specified in the chemical formula of the participating species, and the \index{Hamiltonian}Hamiltonian ${\cal{H}}$ was that for classical nuclear motion on a \textit{\index{Potential Energy Surface}Potential Energy Surface}; this dynamics was assumed to give rise to a critical surface which was such that reaction trajectories cross the surface precisely \textit{once}. The classical nature of the formalism was quite clear because the \index{Uncertainty Principle}Uncertainty Principle precludes the precise specification of position on the critical surface simultaneously with the momentum of the nuclei.}. It is important to note that nowhere did \index{Marcelin}Marcelin elaborate on how the canonical variables were to be chosen, nor even how $n$ could be fixed in any given case. The words `atom', `electron', `nucleus' do not appear anywhere in his thesis, in which respect he seems to have followed the scientific philosophy of his countryman \index{Duhem}Duhem \cite{PD1}. On other pages in the thesis \index{Marcelin}Marcelin refered to the `structure' (also `architecture') of a molecule and to molecular `oscillations' but never otherwise invoked the atomic structural conception of a molecule due to e.g. van 't Hoff, although he was very well aware of van 't Hoff's Physical Chemistry.

Contemporary with \index{Marcelin}Marcelin's investigation of chemical reaction rates was the introduction of a completely novel model of an atom due to Rutherford. However it quickly became apparent that Rutherford's  solar system model of the atom (planetary electrons moving about a central nucleus) cannot avoid eventual collapse if classical \index{electrodynamics}electrodynamics applies to it. This is because of \index{Earnshaw}Earnshaw's theorem which states that it is impossible for a collection of charged particles to maintain a static equilibrium purely through electrostatic  forces \cite{SE:42}. This is the classical result that \index{Bohr}Bohr alluded to in his 1922 Nobel lecture \cite{NB2} to rule out 
an electrostatical explanation for the stability of atoms and molecules. 

The theorem may be proved by demonstrating a contradiction. Suppose the charges are at rest and consider the motion of a particular charge $e_n$ 
in the electric field, ${\bf E}$, generated by all of the other charged
particles. Assume that this particular charge has $e_n~>~0$. The 
equilibrium position of this particle is the point 
${\bf x}_{n}^{0}$ where ${\bf E}({\bf x}_{n}^{0})={\bf 0}$, 
since the force on the charge is $e_n {\bf E}({\bf x}_{n})$ 
(the Lorentz force for this static case). Obviously, 
${\bf x}_{n}^{0}$  cannot be the equilibrium position of any other 
particle. However, in order for ${\bf x}_{n}^{0}$ to be 
a stable equilibrium point, the particle must experience a restoring 
force when it is displaced from ${\bf x}_{n}^{0}$ in 
any direction. For a positively charged particle at ${\bf x}_{n}^{0}$, 
this requires that the electric field points radially towards 
${\bf x}_{n}^{0}$ at all neighbouring points. But from Gauss's law applied 
to a small sphere centred on  ${\bf x}_{n}^{0}$, this corresponds to 
a \textit{negative} flux of ${\bf E}$ through the surface of the 
sphere, implying the presence of a \textit{negative charge} 
at ${\bf x}_{n}^{0}$, contrary to our original assumption. Thus 
${\bf E}$ cannot point radially towards ${\bf x}_{n}^{0}$ at all 
neighbouring points, that is, there must be some neighbouring points 
at which ${\bf E}$ is directed away from ${\bf x}_{n}^{0}$. Hence, 
a positively charged particle placed at ${\bf x}_{n}^{0}$ will always  
move towards such points. There is therefore no static 
\index{equilibrium configuration}equilibrium configuration. According to classical \index{electrodynamics}electrodynamics accelerated charges must radiate electromagnetic energy, and hence lose \index{kinetic energy}kinetic energy, so even a dynamical model cannot be stable according to purely classical theory.

Molecular models which can be represented in terms of the (\index{phase-space}phase-space) variables of \index{classical dynamics}classical dynamics had a far-reaching influence on the interpretation of molecular spectra after the dissemination of \index{Bohr}Bohr's quantum theory of atoms and molecules based on transitions between \index{stationary state}stationary states \cite{NB1}. An important feature of his new theory was that classical \index{electrodynamics}electrodynamics should be deemed to be still operative when transitions took place, but \textit{not} when the system was in a \index{stationary state}stationary state, by fiat. \index{Bohr}Bohr had originally used the fact that two particles with \index{Coulomb}Coulombic interaction lead to a \index{Hamiltonian}Hamiltonian problem that is completely soluble by separation of variables. With more particles and \index{Coulomb}Coulombic interactions this is no longer true; however by largely qualitative reasoning he was able to develop a quantum theory of the atom and the Periodic Table (reviewed in \cite{NB2}). Furthermore by the introduction of Planck's constant $h$ through the angular momentum quantization condition, \index{Bohr}Bohr solved another problem of the classical theory. In classical \index{electrodynamics}electrodynamics the only characteristic length available is the classical radius $r_o$ for a charged particle. This is obtained by equating the rest-mass energy for the charge to the electrostatic energy of a charged sphere of radius $r_o$.
\begin{displaymath}
r_o~=~\left(\frac{e^2}{4\pi\epsilon_0 mc^2}\right)
\end{displaymath}
For an electron this yields $r_o~\approx 2.8 \times 10^{-15}$m and an even smaller value for any nucleus. It was clear that this was far too small to be relevant to an atomic theory; of course the \index{Bohr}Bohr radius $a_o~\approx~ 0.5 \times 10^{-10}$m is of just the right dimension.

\index{Bohr}Bohr's theory developed into the \index{Old Quantum Theory}Old Quantum Theory which was based on a \index{phase-space}phase-space description of an atomic-molecular system and theoretical techniques originally developed in celestial mechanics. These came from the application of the developing quantum theory to molecular band spectra by Schwarzschild \cite{KS} and Heurlinger \cite{TH} who used it to describe the quantized vibrational and rotational energies of small molecules (diatomic and symmetric top structures). Schwarzschild, an astrophysicist, was responsible for the introduction of action-angle methods as a basis for quantization in atomic/molecular theory. Heurlinger assumed a quantization of the energy of the nuclear vibration analogous to that used by Planck for his ideal linear oscillators, with the possibility of anharmonic behaviour. Thus a force-law or \textit{potential energy} depending on the separation of the nuclei, for a given arrangement of the electrons, was required.

The basic calculational tool was a perturbation theory approach developed enthusiastically by \index{Born}Born \cite{MB} and \index{Sommerfeld}Sommerfeld \cite{AS} with their research assistants. The solution of the \index{Hamiltonian}Hamiltonian equations of motion could be attempted via the Hamilton-Jacobi method based on canonical transformations of the action-angle variables. This leads to an expression for the energy that is a function of the action integrals only. The action (or `phase') integrals are constants of the motion, and are also \index{adiabatic}adiabatic invariants \cite{PE}, and as such are natural objects for quantization according to the `quantum conditions'. Thus for a separable system with $k$ degrees of freedom and action integrals \{$J_i,~i=1\ldots j\leq k$\}, the quantum conditions according to \index{Sommerfeld}Sommerfeld are
\begin{equation}
J_i~\equiv~\oint p_i~dq_i~=~n_i h,~~~~i~=~1,\ldots j
\label{action}
\end{equation}
where the $n_i$ are non-negative integers ($j < k$ in case of degeneracy). Here it is assumed that each $p_i$ is a periodic function of only its corresponding conjugate coordinate $q_i$, and the integration is taken over a period of $q_i$. An important principle, due to \index{Bohr}Bohr, was that slow, continuous (`\index{adiabatic}adiabatic') deformations of an atomic system kept the system in a \index{stationary state}stationary state \cite{NB3}, \cite{NB4}. Thus the action integrals for a \index{Hamiltonian}Hamiltonian depending on parameters that vary slowly in time are conserved under slow changes of the parameters\footnote{This is strictly true only for integrable \index{Hamiltonian}Hamiltonians \cite{MBe}.}. This could be applied to the problem of chemical bonding by treating the nuclear positions as the slowly varying parameters in an \index{adiabatic}adiabatic transformation of the \index{Hamiltonian}Hamiltonian for the electrons in the presence of the nuclei.

We now know that systems of more than 2 particles with \index{Coulomb}Coulomb interactions may have very complicated dynamics; \index{Newton}Newton famously struggled to account quantitatively for the orbit of the moon in the earth-moon-sun problem ($n=3$). The underlying reason for his difficulties is the existence of solutions carrying the signature of chaos \cite{MG} and this implies that there are classical trajectories to which the quantum conditions simply cannot be applied\footnote{The difficulties for action-angle quantization posed by the existence of chaotic motions in non-separable systems \cite{HP} were recognized by Einstein at the time the \index{Old Quantum Theory}Old Quantum Theory was developed \cite{AE}.} because the integrals in (\ref{action}) do not exist \cite{ICP}. We also know that the $r^{-1}$ singularity in the classical potential energy can lead to pathological dynamics in which a particle is neither confined to a bounded region, nor escapes to infinity for good. If the two-body interaction $V({\bf
r})$ has a Fourier transform $v({\bf k})$ the total potential energy
can be expressed as
\begin{eqnarray*}
U~&=&~\sum_{i<j}^{n}e_ie_j~V(|{\bf x}_i-{\bf
x}_j|)\\~&=&~-~\frac{n}{2}V(0)~+~\frac{1}{(2\pi)^3}\int d^3{\bf k}~
v({\bf k})\left|\sum_i e_i~ e^{i{\bf k}.{\bf x}_i}\right|^2
\end{eqnarray*}
In the case of the \index{Coulomb}Coulomb interaction $v({\bf k})=4\pi/k^2>0$ and
so the potential energy $U$ is bounded from below by $-nV(0)/2$;
unfortunately for point charges as $r\rightarrow 0, V(r)\rightarrow
\pm\infty$ and collapse may ensue \cite{Th}. 

Attempts were made by \index{Born}Born and his assistants to discuss the \index{stationary state}stationary state energy levels of `simple' non-trivial systems such as He, H$_2^{+}$, H$_2$, H$_2$O. The molecular species were tackled as problems in \textit{\index{electronic structure}electronic structure}, that is, as requiring the calculation of the energy levels for the electron(s) in the field of \textit{fixed} nuclei as a calculation separate from the rotation-vibration of the molecule as a whole. Pauli gave a lengthy qualitative discussion of the possible \index{Bohr}Bohr orbits for the single electron moving in the field of two fixed protons in H$_2^{+}$ but could not obtain the correct \index{stationary state}stationary states \cite{WP}. \index{Nordheim}Nordheim investigated the forces between two hydrogen atoms as they approach each other \index{adiabatic}adiabatically\footnote{This is the earliest reference we know of where the idea of \textit{\index{adiabatic}adiabatic} separation of the electrons and the nuclei is proposed explicitly.} in various orientations consistent with the quantum conditions. Before the atoms get close enough for the attractive and repulsive forces to balance out, a sudden discontinuous change in the electron orbits takes place and the electrons cease to revolve solely round their parent nuclei. \index{Nordheim}Nordheim was unable to find an interatomic distance at which the energy of the combined system was less than that of the separated atoms; this led to the conclusion that the use of classical mechanics to discuss the \index{stationary state}stationary states of the molecular electrons had broken down comprehensively \cite{LN},\cite{ECK}. This negative result was true of all the molecular calculations attempted within the \index{Old Quantum Theory}Old Quantum Theory framework which was simply incapable of accounting for covalent bonding \cite{HK}.

The most ambitious application of the \index{Old Quantum Theory}Old Quantum Theory to molecular theory was made by \index{Born}Born and \index{Heisenberg}Heisenberg \cite{BH}. They started from the usual non-relativistic \index{Hamiltonian}Hamiltonian (\ref{molHam}) for a system comprised of $n$ electrons and $N$ nuclei interacting via \index{Coulomb}Coulombic forces. They assumed there is an arrangement of the nuclei which is a stable equilibrium, and use that (a \index{molecular structure}molecular structure) as a reference configuration for the calculation. Formally the rotational motion of the system can be dealt with by requiring the coordinates for the reference structure to satisfy\footnote{This also deals with the uninteresting overall translation of the molecule.} what were later to become known as `the \index{Eckart}Eckart conditions' \cite{CE}. Then with a suitable set of internal variables and
\begin{displaymath}
\lambda~=~\left(\frac{m}{M}\right)^{\frac{1}{2}}
\end{displaymath}
as the expansion parameter, the \index{Hamiltonian}Hamiltonian was expressed as a series
\begin{equation}
H~=~H_o~+~\lambda^2H_2~+~\ldots
\label{BHexpans}
\end{equation}
to be treated by the action-angle perturbation theory \index{Born}Born had developed.
The `unperturbed' \index{Hamiltonian}Hamiltonian $H_o$ is the full \index{Hamiltonian}Hamiltonian for the electrons with the nuclei fixed at the equilibrium structure, $H_2$ is quadratic in the nuclear variables (harmonic oscillators) and also contains the rotational energy\footnote{The rotational and vibrational energies occur together because of the choice of the parameter $\lambda$; as is well-known, \index{Born}Born and \index{Oppenheimer}Oppenheimer later showed that a better choice is to take the quarter power of the mass ratio as this separates the vibrational and rotational energies in the orders of the perturbation expansion \cite{BO}.}, while $\ldots$ stands for higher order anharmonic vibrational terms. $H_1$ may be dropped because of the equilibrium condition. With considerable effort there follows the usual separation of molecular energies, although of course no concrete calculation was possible within the \index{Old Quantum Theory}Old Quantum Theory framework. It is noteworthy that their calculation gives the electronic energies at a \textit{single} configuration because the perturbation calculation requires the introduction of the (assumed) equilibrium structure.  This is different from the \textit{\index{adiabatic}adiabatic} approach \index{Nordheim}Nordheim tried (unsuccessfully) to get the electronic energy at \textit{any} separation of the nuclei \cite{LN}.

\section{Quantum Theory}
\label{quant}
With the completion of quantum mechanics in 1925-26, the old problems in atomic and molecular theory were reconsidered and considerable success was achieved. The idea that the dynamics of the electrons and the nuclei should be treated to some extent as separate problems was generally accepted. Thus the \index{electronic structure}electronic structure calculations of \index{London}London \cite{HL} - \cite{FL2} can be seen as a successful reformulation of the approach \index{Nordheim}Nordheim had tried in terms of the older quantum theory, and the idea of `\index{adiabatic}adiabatic separation' is often said to originate in this work. It is however also implied in the closing section of \index{Slater}Slater's early He atom paper where he sketches (but does not carry through) a perturbation method of approximate calculation for molecules in which the nuclei are first held fixed, and the resulting electronic eigenvalue(s) then act as the potential energy for the nuclei \cite{JCS}. A quantum mechanical proof of \index{Ehrenfest}Ehrenfest's \index{adiabatic}adiabatic theorem for time-dependent perturbations was given by \index{Born}Born and \index{Fock}Fock \cite{BF:28}. Most famously though, the quantum mechanical basis for the idea of electronic \index{Potential Energy Surface}Potential Energy Surfaces is commonly attributed to \index{Born}Born and \index{Oppenheimer}Oppenheimer, and it is to a consideration of their famous paper \cite{ BO}  that we now turn.
\subsection{\index{Born}Born and \index{Oppenheimer}Oppenheimer's \textit{Quantum Theory of Molecules}}
\label{BOtheory}
Much of the groundwork for \index{Born}Born and \index{Oppenheimer}Oppenheimer's treatment of the energy levels of molecules was laid down in the earlier attempt by \index{Born}Born and \index{Heisenberg}Heisenberg \cite{BH}. The basic idea of both calculations is that the low-lying excitation spectrum of a molecule can be obtained by regarding the nuclear \index{kinetic energy}kinetic energy as a `small' perturbation of the energy of the electrons for \index{stationary nuclei}stationary nuclei \textit{in an \index{equilibrium configuration}equilibrium configuration}. The physical basis for the idea is the large disparity between the mass of the electron and the masses of the nuclei; classically the light electrons undergo motions on a `fast' timescale ($\tau_e ~\approx~ 10^{-16}~\rightarrow~ 10^{-15}$s), while the vibration-rotation dynamics of the much heavier nuclei are characterized by  `slow' timescales ($\tau_N~\approx~10^{-14}~\rightarrow~10^{-12}$s).   

Consider a system of electrons and nuclei and denote the properties of the former by lower-case letters (mass $m$, coordinates $x$, momenta $p$) and of the latter by capital letters (mass $M$, coordinates $X$, momenta $P$). The small parameter for the perturbation expansion must clearly be some power of $m/M_o$, where $M_o$ can be taken as any one of the nuclear masses or their average. In contrast to the earlier calculation they found the correct choice is
\begin{displaymath}
\kappa~=~\left(\frac{m}{M_o}\right)^{\frac{1}{4}}
\end{displaymath}
rather than \index{Born}Born and \index{Heisenberg}Heisenberg's $\lambda~=\kappa^2$. In an obvious shorthand notation using a coordinate representation the \index{kinetic energy}kinetic energy of the electrons is then\footnote{The details can be found in the original paper \cite{BO}, and in various English language presentations, for example \cite{BHu} - \cite{BOBl}.\label{fnBO}}
\begin{displaymath}
T_e~=~T_e\left(\frac{\partial}{\partial x}\right)
\end{displaymath}
while the nuclear \index{kinetic energy}kinetic energy depends on $\kappa$
\begin{displaymath}
T_N~=~\kappa^4 H_1\left(\frac{\partial}{\partial X}\right)
\end{displaymath}
The \index{Coulomb}Coulomb energy is simply $U(x,X)$. They then define the `unperturbed' \index{Hamiltonian}Hamiltonian 
\begin{equation}
T_e~+~U~=~H_o\left(x,\frac{\partial}{\partial x},X\right)
\label{unpert}
\end{equation}
and express the total \index{Hamiltonian}Hamiltonian as
\begin{equation}
H~=~H_o~+~\kappa^4H_1
\label{fullH}
\end{equation}
with \index{Schr\"{o}dinger}Schr\"{o}dinger equation
\begin{equation}
\bigg(H~-~E\bigg)\psi(x,X)~=~0.
\label{Scheq}
\end{equation} 

At this point in their argument they state

\begin{quote}
Setzt man in (12) [equation(\ref{Scheq}) above] $\kappa~=~0$, so bekommt man eine Differentialgleichung f\"{u}r die $x_k$ allein, in der die $X_l$ als Parameter vorkommen:
\begin{displaymath}
\left\{H_o\left(x,\frac{\partial}{\partial x};X\right)~-~W\right\}\psi~=~0.
\end{displaymath}
Sie stellt offenbar die Bewegung der Elektronen bei festgehaltenen Kernen dar\footnote{If one sets $\kappa=0$....one obtains a differential equation in the $x_k$ alone, the $X_l$ appearing as parameters:.....Evidently, this represents the electronic motion for \index{stationary nuclei}stationary nuclei.}.
\end{quote}

This splitting of the \index{Hamiltonian}Hamiltonian into an `unperturbed' part ($\kappa=0$) and a `perturbation' is essentially the same as in the earlier \index{Old Quantum Theory}Old Quantum Theory version \cite{BH}. The difference here is that the action-angle perturbation theory of the \index{Old Quantum Theory}Old Quantum Theory is replaced by \index{Schr\"{o}dinger}Schr\"{o}dinger's quantum mechanical perturbation theory. In the following it is understood that the overall translational motion of the molecule has been separated off by a suitable coordinate transformation; this is always possible. The initial step in setting up the perturbation expansion involves rewriting the \index{Hamiltonian}Hamiltonian $H_o$ as a series in increasing powers of $\kappa$. This is achieved by introducing new relative coordinates that depend on $\kappa$
\begin{equation}
X~=~X_o~+~\kappa \zeta
\label{eqcoord}
\end{equation}
for some fixed $X_o$, and using the \{$\zeta$\} as the nuclear variables, in an expansion of $H_o$ about $X_o$. 

Then as usual the eigenfunction and eigenvalue of (\ref{Scheq}) are presented as series in $\kappa$
\begin{eqnarray*}
\psi~&=&~\psi^{(0)}~+~\kappa~\psi^{(1)}~+~\kappa^2~\psi^{(2)}~+\ldots\\
E~&=&~E^{(0)}~+~\kappa~E^{(1)}~+~\kappa^2~E^{(2)}~+\ldots,
\end{eqnarray*}
the expansions are substituted into the \index{Schr\"{o}dinger}Schr\"{o}dinger equation 
(\ref{Scheq}), and the terms separated by powers of $\kappa$. This gives a set of equations to be solved sequentially. The crucial observation that makes the calculation successful is the choice of $X_o$; the \index{Schr\"{o}dinger}Schr\"{o}dinger equation for the unperturbed \index{Hamiltonian}Hamiltonian $H_o$ can be solved for any choice of the nuclear parameters $X$, and yields\footnote{$W(X)$ in the notation of the above quotation.} an unperturbed energy $E(X)$ for the configuration $X$. For the consistency of the whole scheme however it turns out (cf footnote \ref{fnBO}) that $X_o$ in equation (\ref{eqcoord}) cannot be arbitrarily chosen, but must correspond to a \textit{minimum} of the electronic energy. That there is such a point is assumed to be self-evident for the case of a stable molecule. The result of the calculation was a triumph; the low-lying energy levels of a stable molecule can be written in the form
\begin{equation}
E_{\mbox{Mol}}~=~E_{\mbox{Elec}}~+~\kappa^2~E_{\mbox{Vib}}~+~\kappa^4~E_{\mbox{Rot}}~+\ldots
\label{Emol}
\end{equation}
in agreement with a considerable body of spectroscopic evidence. The eigenfunctions that correspond to these energy levels are simple products of an electronic wavefunction obtained for the equilibrium geometry and suitable vibration-rotation wavefunctions for the nuclei. 

The \index{Born}Born and \index{Heisenberg}Heisenberg calculation \cite{BH} had been performed while \index{Heisenberg}Heisenberg was a student with \index{Born}Born; \index{Kragh}Kragh \cite{HK} quotes \index{Heisenberg}Heisenberg's later view of it in the following terms
\begin{quote}
As an exasperated \index{Heisenberg}Heisenberg wrote to Pauli, ``The work on molecules I did with \index{Born}Born.....contains bracket symbols [Klammersymbole] with up to 8 indices and will probably be read by no one.'' Certainly, it was not read by the chemists.
\end{quote}
Curiously that may have initially been the fate of \index{Born}Born and \index{Oppenheimer}Oppenheimer's paper. As noted by one of us many years ago, a survey of the literature up to about 1935 shows that the paper was hardly if ever mentioned, and when it was mentioned, its arguments were used as \textit{a posteriori} justification for what was being done anyway \cite{BTS}. What was being done was the general use in molecular spectroscopy and chemical reaction theory of the idea of \index{Potential Energy Surface}Potential Energy Surfaces on which the nuclei moved. As we have seen, that idea is \textit{not} to be found in the approach taken by \index{Born}Born and \index{Oppenheimer}Oppenheimer which used (and had to use) a single privileged \textit{point} in the nuclear configuration space - the assumed equilibrium arrangement of the nuclei \cite{BO}.

In 1935 a significant event was the publication of the famous textbook \textit{Introduction to Quantum Mechanics} \cite{PW} which was probably the first textbook concerned with quantum mechanics that addressed in detail problems of interest to chemists. Generations of chemists and physicists took their first steps in quantum theory with this book, which is still available in reprint form. Chapter X of the book is entitled \textit{The Rotation and Vibration of Molecules}; it starts by summarizing the empirical results of molecular spectroscopy which are consistent with (\ref{Emol}). The authors then turn to the wave equation for a general collection of electrons and nuclei and remark that its \index{Schr\"{o}dinger}Schr\"{o}dinger wave equation may be solved approximately by a procedure that they attribute to \index{Born}Born and \index{Oppenheimer}Oppenheimer; first solve the wave equation for the electrons alone, with the nuclei in a fixed configuration, and then solve the wave equation for the nuclei alone, in which a characteristic energy value [eigenvalue] of the electronic wave equation, regarded as a function of the internuclear distances, occurs as a potential function. After some remarks about the coordinates they say
\begin{quote}
The first step in the treatment of a molecule is to solve this electronic wave equation for all configurations of the nuclei. It is found that the characteristic values $U_n(\xi)$ of the electronic energy are continuous functions of the nuclear coordinates $\xi$. For example, for a free diatomic
molecule the electronic energy function for the most stable electronic state ($n~=~0$) is a function only of the distance $r$ between the two nuclei, and it is a continuous function of $r$, such as shown in Figure 34-2.
\end{quote}
Figure 34-2 referred to here is a Morse potential function. Later in the book where they give a brief introduction to activation energies of chemical reactions they explicitly cite \index{London}London \cite{FL2} as the origin of the idea of \index{adiabatic}adiabatic nuclear motion on a \index{Potential Energy Surface}Potential Energy Surface, though there is also a nod back towards Chapter X. Although it is now almost universal practice to refer to treating the nuclei as classical particles that give rise to an electronic energy surface as `making the \index{Born}Born-\index{Oppenheimer}Oppenheimer approximation' it is our opinion that the justification for such a strategy is not to be found in \textit{The Quantum Theory of Molecules}, \cite{BO}. Nor is it to be found in the early papers of \index{London}London \cite{HL} - \cite{FL2} where it was simply assumed as a reasonable thing to do. And it is certainly the case that \index{Born}Born and \index{Oppenheimer}Oppenheimer did \textit{not} show the electronic energy to be a continuous function of the nuclear coordinates; that was first demonstrated for a diatomic molecule forty years after \index{Pauling}Pauling and \index{Wilson}Wilson's book was published (see \S \ref{approx}).

\subsection{\index{Born}Born and the elimination of electronic motion}
\label{Brnadi}
Many years after his work with \index{Heisenberg}Heisenberg and \index{Oppenheimer}Oppenheimer, \index{Born}Born returned to the subject of molecular quantum theory and developed a different account of the separation of electronic and nuclear motion \cite{BHu}, \cite{Bmol}. It is to this method that the expression `\index{Born}Born-\index{Oppenheimer}Oppenheimer approximation' usually refers in modern work. Consider the unperturbed electronic \index{Hamiltonian}Hamiltonian $H_o(x,X_f)$ at a fixed nuclear configuration $X_f$ that corresponds to some \index{molecular structure}molecular structure (not necessarily an equilibrium structure). The \index{Schr\"{o}dinger}Schr\"{o}dinger equation for $H_o$ is
\begin{equation}
\bigg(H_o(x,X_f)~-~E^{o}(X_f)_m\bigg)\varphi(x,X_f)_m~=~0.
\label{scheqcn}
\end{equation}
This equation can have both bound-state and continuum eigenfunctions; the \textit{bound-state} eigenvalues considered as functions of the $X_f$ are the molecular \index{Potential Energy Surface}Potential Energy Surfaces. \index{Born}Born proposed to solve the full molecular \index{Schr\"{o}dinger}Schr\"{o}dinger equation, (\ref{Scheq}) by an expansion
\begin{equation}
\psi(x,X)~=~\sum_m \Phi(X)_m~\varphi(x,X)_m
\label{wfnexpan}
\end{equation}
with coefficients \{$\Phi(X)_m$\} that play the role of nuclear
wavefunctions. As in the original calculation (\S \ref{BOtheory}) a crucial step is to assign the nuclear coordinates the role of parameters in the \index{Schr\"{o}dinger}Schr\"{o}dinger equation (\ref{scheqcn}) for the electronic \index{Hamiltonian}Hamiltonian; it differs from the earlier approach of \index{Born}Born and \index{Oppenheimer}Oppenheimer because now the values of $X_f$ range over the whole nuclear configuration space. Substituting this expansion into (\ref{Scheq}), multiplying the result by $\varphi(x,X)_n^{*}$ and integrating over the electronic coordinates $x$ leads to an infinite dimensional system of coupled equations for the nuclear functions
\{$\Phi$\},
\begin{equation}
\big(T_N~+~E^{o}(X)_n~-~E\big)\Phi(X)_n~+~\sum_{nn'}C(X,P)_{nn'}\Phi(X)_{n'}~=~0
\label{coupl}
\end{equation}
where the coupling coefficients \{$C(X,P)_{nn'}$\} have a well-known
form which we need not record here \cite{BHu}.  

In this formulation the \index{adiabatic}adiabatic approximation consists of retaining only the diagonal terms in the coupling matrix ${\bf C}(X,P)$, for then a state function can be written as
\begin{equation}
\psi(x,X)~\approx~\psi(x,X)_{n}^{\mbox{AD}}~=~\varphi(x,X)_n~\Phi(X)_n.
\label{adibat}
\end{equation}
and a product wavefunction corresponds to additive electronic and nuclear energies. The special character of the electronic wavefunctions \{$\varphi(x,X)_m$\} is, by (\ref{scheqcn}), that they diagonalize the electronic \index{Hamiltonian}Hamiltonian $H_o$; they are said to define an `\index{adiabatic}adiabatic' basis (cf the approximate form (\ref{adibat})) because the electronic state label $n$ is not altered as $X$ varies. The \index{Born}Born approach does not really require the diagonalization of $H_o$; it is perfectly possible to define other representations of the electronic expansion functions through unitary transformations of the \{$\varphi$\}, with concomitant modification of the coupling matrix ${\bf C}$. This leads to so-called `diabatic' bases; the freedom to choose the representation is very important in practical applications to spectroscopy and atomic/molecular collisions \cite{TFOM:71,GGH:87}.

\subsection{Formal quantum theory of the molecular \index{Hamiltonian}Hamiltonian}
\label{FQT}
We now start again and develop the quantum theory of the \index{Hamiltonian}Hamiltonian for a collection of $n$ charged particles with \index{Coulomb}Coulombic interactions\footnote{The reader may find it helpful to refer to the Appendix which summarizes some mathematical notions that are needed here, and illustrates them in a simple model of coupled oscillators with two degrees of freedom.}. We remind ourselves again from \S \ref{Intro} that for particles with classical \index{Hamiltonian}Hamiltonian variables \{${\bf q}_i,{\bf p}_i$\} this is
\begin{equation}
\mathsf{H}~=~\sum_i^n\frac{p_i^{2}}{2m_i}~+~\sum_{i < j}^{n}\frac{e_i e_j}{4
\pi \epsilon_0 |{\bf q}_i~-~{\bf q}_j|}
\label{ClassHam}
\end{equation}
with the non-zero Poisson-bracket
\begin{displaymath}
\{{\bf x}_i,{\bf p}_j\}~=~\delta_{ij}
\end{displaymath}
Let us denote the classical dynamical variables for the electrons collectively as ${\bf x}, {\bf p}$, and those for the nuclei by ${\bf X}, {\bf P}$ and denote the classical \index{Hamiltonian}Hamiltonian by $\mathsf{H}({\bf x},{\bf p},{\bf X},{\bf P})$. After the customary canonical quantization these variables become time-independent operators in a \index{Schr\"{o}dinger}Schr\"{o}dinger representation 
\begin{displaymath}
{\bf x}~\rightarrow~\hat{\bf x}~etc.
\end{displaymath}
In the following it will be important to distinguish between operators and c-numbers, so in the following we will use the $\hat{\bf x}$ notation for operators, and make no special choice of representation.

As we have seen, the idea that the \index{kinetic energy}kinetic energy of the massive nuclei could be treated as a perturbation of the electronic motion was first formulated in the framework of the \index{Old Quantum Theory}Old Quantum Theory. The idea was to  separate the classical \index{Hamiltonian}Hamiltonian $\mathsf{H}$ into two parts to isolate the nuclear momentum variables
\begin{equation}
\mathsf{H}({\bf x},{\bf p},{\bf X},{\bf P})~=~H_o({\bf x},{\bf p},{\bf X})~+~\kappa^4~H_1({\bf P}).
\label{OQTHam}
\end{equation}
According to Hamilton's equations for the unperturbed problem
\begin{equation}
\frac{d{\bf X}}{dt}~=~\{{\bf X},H_o\}~=~0,
\label{cleqnmot}
\end{equation}
using Poisson-bracket notation, which was interpreted (correctly) as describing the dynamics of the electrons in the field of \index{stationary nuclei}stationary nuclei. This was the starting point of \index{Born}Born and \index{Heisenberg}Heisenberg's calculations \cite{BH}.

Let us now move to quantum theory and recast (\ref{OQTHam}) as an operator relation, writing the molecular \index{Hamiltonian}Hamiltonian operator as
\begin{equation}
\mathsf{\hat{H}}(\hat{\bf x},\hat{\bf p},\hat{\bf X},\hat{\bf P})~=~\hat{H}_o(\hat{\bf x},\hat{\bf p},\hat{\bf X})~+~\kappa^4~\hat{H}_1(\hat{\bf P})
\label{QMHam}
\end{equation}
with equation of motion under $\hat{H}_o$
\begin{equation}
i\hbar\frac{d\hat{\bf X}}{dt}~=~[\hat{\bf X},\hat{H}_o]~=~0
\label{qmeqnmot}
\end{equation}
from which we infer the nuclear \index{position operator}position operators $\hat{\bf X}$ are constants of the motion under $\hat{H}_o$. We no longer make the
interpretation that follows from (\ref{cleqnmot}) since specifying precisely the positions \{{\bf X}\} for \textit{stationary} nuclei violates the \index{Uncertainty Principle}Uncertainty Principle. Instead (\ref{qmeqnmot}) leads to a completely different conclusion (see below).

We must now take a little bit of care about the definition of the variables, and dispose of the uninteresting overall motion of the molecule \cite{SW:12}. Since the \index{Coulomb}Coulomb interaction only depends on interparticle distances it is translation invariant, and therefore the total \index{momentum operator}momentum operator $\mathsf{\hat{P}}$
\begin{displaymath}
\mathsf{\hat{P}}~=~\sum_{n}{\hat{\bf p}}_n
\end{displaymath}
commutes with $\mathsf{\hat{H}}$. It follows that the molecular \index{Hamiltonian}Hamiltonian may be written as a \index{direct integral}direct integral
\begin{equation}
\mathsf{\hat{H}}~=~\int_{\mbox{\bf R}^3}^{\oplus}\hat{H}(P)~dP
\label{dirint}
\end{equation}
where \cite{LMS}
\begin{equation}
\hat{H}(P)=\frac{{P}^2}{2M_T} +\mathsf{\hat{H}}'
\label{sep}
\end{equation}
is the \index{Hamiltonian}Hamiltonian at fixed total momentum $P$ and $M_T$ is the molecular mass. The internal \index{Hamiltonian}Hamiltonian $\mathsf{\hat{H}}'$ is independent of the centre-of-mass variables and is explicitly translation invariant. The form of $\mathsf{\hat{H}}'$ is not uniquely fixed but whatever coordinates are chosen the essential point is that it is always the same operator specified in (\ref{sep}) acting on a \index{Hilbert space}Hilbert space $\mathfrak{H}$ that may be parameterized by functions of the electron and nuclear coordinates.

The separation of the centre-of-mass terms from the internal \index{Hamiltonian}Hamiltonian is the same in quantum mechanics as in classical mechanics so we need not distinguish operators from classical variables in this step. It is convenient to choose the centre-of-nuclear mass for the definition of suitable internal coordinates\footnote{It is always possible to split off the \index{kinetic energy}kinetic energy of the centre-of-mass without any approximation; with this choice we retain the separation of the electronic and nuclear kinetic energies as well, as in (\ref{Hdecomp}). Explicit formulae are given in e.g. \cite{WS:03} where it is shown that the nuclear \index{kinetic energy}kinetic energy terms involve reciprocals of the nuclear masses, so that overall, the nuclear \index{kinetic energy}kinetic energy is proportional to $\kappa^4$.\label{fn12}}. Let ${\bf t}^{\rm e}$ be a set of internal electronic coordinates defined as the original electronic coordinates ${\bf x}$ referred to the centre-of-nuclear mass, and let ${\bf t}^{\rm n}$ be a set of internal nuclear coordinates constructed purely from the original nuclear coordinates ${\bf X}$. If there are $s$ electrons and $M$ nuclei, there are $s$ internal electronic coordinates, and $M-1$ internal nuclear coordinates.  There are corresponding canonically conjugate internal momentum variables. In terms of these variables the total \index{kinetic energy}kinetic energy of all the particles can be decomposed into the form
\begin{equation}
\mathsf{T}_0~=~\mathsf{T}_{\mbox{CM}}~+~\mathsf{T}_{\mbox{N}}~+~\mathsf{T}_{\mbox{e}}
\label{Hdecomp}
\end{equation}
where $\mathsf{T}_{\mbox{CM}}$ is the \index{kinetic energy}kinetic energy for the centre-of-mass, $\mathsf{T}_{\mbox{N}}$ is the \index{kinetic energy}kinetic energy for the nuclei expressed purely in terms of the internal nuclear momentum variables, and $\mathsf{T}_{\mbox{e}}$ is the \index{kinetic energy}kinetic energy for the electrons expressed purely in terms of the internal electronic momentum variables. The \index{Coulomb}Coulomb energy can be expressed purely in terms of the internal coordinates, $U~=~U({\bf t}^{\rm e},{\bf t}^{\rm n})$. These relations are true both classically and in quantum mechanics with a suitable operator interpretation.

In parallel with the decomposition in equation (\ref{OQTHam}), we define the quantum mechanical `electronic' \index{Hamiltonian}Hamiltonian as
\begin{equation}
\mathsf{\hat{H}}^{\mbox{elec}}~=~ \mathsf{\hat{T}}_{\mbox{e}}~+~\hat{U}(\hat{\bf t}^{\rm e},\hat{\bf t}^{\rm n})
\label{elecHam}
\end{equation}
so that after dropping the uninteresting \index{kinetic energy}kinetic energy for the overall centre-of-mass, we see that the internal \index{Hamiltonian}Hamiltonian has the form,
\begin{equation}
\mathsf{\hat{H}}'~=~\mathsf{\hat{H}}^{\mbox{elec}}~+~\mathsf{\hat{T}}_{\mbox{N}}
\label{intHam}
\end{equation}
where, as before, the nuclear \index{kinetic energy}kinetic energy term is proportional to $\kappa^4$ (see footnote \ref{fn12}). Its \index{Schr\"{o}dinger}Schr\"{o}dinger equation may be written
\begin{equation}
\mathsf{\hat{H}}'~|\Psi_m\rangle~=~E_m~|\Psi_m\rangle
\label{intScheq}
\end{equation}
where $m$ is used to denote a set of quantum numbers ($J~M~p~r~i$): $J$ and $M$ for the angular momentum state: $p$ specifying the parity of the state:  $r$ specifying the permutationally allowed irreducible representations within the groups of identical particles, and $i$ to specify a particular energy value. Any bound state (a `molecule') has an energy lying below the start of the essential spectrum.

Now just as in equation (\ref{qmeqnmot}) $\mathsf{\hat{H}}^{\mbox{elec}}$ is independent of the nuclear \index{momentum operator}momentum operators and so it commutes with the internal nuclear \index{position operator}position operators
\begin{equation}
[\mathsf{\hat{H}}^{\mbox{elec}}, \mathsf{\hat{\bf t}}^{\rm n}]~=~0.
\label{poscomm}
\end{equation}
They may therefore be simultaneously diagonalized and we use this
property to characterize the \index{Hilbert space}Hilbert space ${\cal{H}}$ for
$\mathsf{\hat{H}}^{\mbox{elec}}$. Let ${\bf b}$ be some eigenvalue of the $\hat{\bf t}^{\rm n}$ corresponding to choices \{${\bf x}_g~=~{\bf a}_g, g = 1,\ldots M$\} in the laboratory-fixed frame; then the \{${\bf a}_g$\} describe a classical nuclear geometry. The set, $X$, of all ${\bf b}$ is ${\bf R}^{3(M-1)}$. We denote the \index{Hamiltonian}Hamiltonian $\mathsf{\hat{H}}^{\mbox{elec}}$ evaluated at the nuclear position eigenvalue ${\bf b}$ as
$\mathsf{\hat{K}}({\bf b},\hat{\bf t}^{\rm e})_o~=~\mathsf{\hat{K}}_o$ for short; this $\mathsf{\hat{K}}_o$ is very like the usual \index{clamped-nuclei} clamped-nuclei \index{Hamiltonian}Hamiltonian but it is explicitly translationally invariant, and has an extra term, which is often called the Hughes-\index{Eckart}Eckart term, or the mass polarization term. Its \index{Schr\"{o}dinger}Schr\"{o}dinger equation is of the same form as (\ref{scheqcn}), with eigenvalues $E^{o}({\bf b})_k$ and corresponding eigenfunctions $\varphi({\bf t}^{\rm e},{\bf b})_k$, 
\begin{equation}
\mathsf{\hat{K}}_o~\varphi({\bf b},{\bf t}^{\rm e})_k~=~E^{o}({\bf b})_k~\varphi({\bf b},{\bf t}^{\rm e})_k
\label{HoSch}
\end{equation}
As before its spectrum in general contains a discrete part below a continuum,
\begin{equation}~
  \sigma({\bf b})~\equiv~\sigma(\mathsf{\hat{K}}({\bf b},\hat{\bf t}^{\rm e})_o)~=~\bigg[E^{o}({\bf b})_0, \ldots
    E^{o}({\bf b})_m\bigg)~\bigcup~ \bigg[\Lambda({\bf b}),\infty\bigg) 
\label{speccn}
\end{equation}

Note that for other than diatomic molecules, it is not possible to proceed further and separate out explicitly the rotational motion. For any choice of ${\bf b}$ the eigenvalues of $\mathsf{\hat{K}}_o$ will depend only upon the shape of the geometrical figure formed by the \{${\bf a}_g$\}, being independent of its orientation. It is possible to introduce a so-called body-fixed frame by transforming to a new coordinate system built out of the ${\bf b}$ consisting of three angular variables and $3M-6$ internal coordinates. In so doing however one cannot avoid angular momentum terms arising which couple the electronic and nuclear variables, and so there is no longer a clean separation of the \index{kinetic energy}kinetic energy into an electronic and a nuclear part. Moreover no \textit{single} specification of body-fixed coordinates can be given that describes \textit{all} possible nuclear configurations.

The internal molecular \index{Hamiltonian}Hamiltonian $\mathsf{\hat{H}}'$ in
(\ref{sep}) and the \index{clamped-nuclei} clamped-nuclei like operator $\mathsf{\hat{K}}_o$ just
defined can be shown to be essentially \index{self-adjoint}self-adjoint (on their respective \index{Hilbert space}Hilbert spaces) by reference to the Kato-Rellich theorem \cite{RS} because
in both cases there are \index{kinetic energy}kinetic energy operators that dominate the
(singular) \index{Coulomb}Coulomb interaction; they therefore have a complete set of
eigenfunctions. As regards $\mathsf{\hat{H}}^{\mbox{elec}}$, we have a family of \index{Hilbert space}Hilbert spaces \{$\cal{H}({\bf b})$\} which are parameterized by the nuclear position vectors ${\bf b} \in X$ that are the `eigenspaces' of the family of \index{self-adjoint}self-adjoint operators $\mathsf{\hat{K}}_o$; from them we can construct a big \index{Hilbert space}Hilbert space as a \index{direct integral}direct integral over all the ${\bf b}$ values 
\begin{equation}
{\cal{H}}~=~\int^{\oplus}_{X}{\cal{H}}({\bf b})~d{\bf b}
\label{bigH}
\end{equation}
and this is the \index{Hilbert space}Hilbert space for $\mathsf{\hat{H}}^{\mbox{elec}}$ in
(\ref{elecHam}).

Equation (\ref{bigH}) leads directly to a fundamental result; since $\mathsf{\hat{H}}^{\mbox{elec}}$ commutes with all
the \{$\mathsf{\hat{\bf t}}^{\rm n}$\}, it  has the \index{direct integral}direct integral
decomposition 
\begin{equation}
\mathsf{\hat{H}}^{\mbox{elec}}~=~  \int^{\oplus}_{X}\mathsf{\hat{K}}({\bf b},\hat{\bf t}^{\rm e})_o~d{\bf b}.
\label{decomp}
\end{equation}
Even if the `\index{clamped-nuclei} clamped-nuclei' \index{Hamiltonian}Hamiltonian has a set of discrete states - \index{Potential Energy Surface}Potential Energy Surfaces - equation (\ref{decomp}) implies that the unperturbed \index{Hamiltonian}Hamiltonian\footnote{After the elimination of the centre-of-mass variables $\mathsf{\hat{H}}^{\mbox{elec}}$ is playing the role of $\hat{H}_o$ in (\ref{QMHam}).}, $\mathsf{\hat{H}}^{\mbox{elec}}$, \textit{has purely \index{continuous spectrum}continuous spectrum} (cf the Appendix),
\begin{displaymath}
\sigma~=~\sigma(\mathsf{\hat{H}}^{\mbox{elec}})~=~\bigcup_{\bf b} ~\sigma({\bf b})~\equiv [V_0,\infty)  
\end{displaymath}
where $V_0$ is the minimum value of $E({\bf b})_0$; in the diatomic
molecule case this is the minimum value of the usual ground-state potential energy curve $E_0(r)$. The operator $\mathsf{\hat{H}}^{\mbox{elec}}$ has no localized eigenfunctions; rather, its eigenfunctions are continuum functions. To avoid any misunderstanding, we emphasize that this result has nothing to do with the \index{continuous spectrum}continuous spectrum of the full molecular \index{Hamiltonian}Hamiltonian associated with the centre-of-mass motion which can be dealt with trivially in the preliminaries.

A possibly helpful way to think about this paradoxical result is as follows. The quantum mechanical molecular \index{Hamiltonian}Hamiltonian for a collection of electrons and nuclei with \index{Coulomb}Coulomb interactions is a function of position and \index{momentum operator}momentum operators for all the specified electrons and all the nuclei.  If now we separate off the terms containing all the nuclear \index{momentum operator}momentum operators (the terms proportional to $\kappa^4$) what is left must be a function of position and \index{momentum operator}momentum operators for the electrons \textit{and \index{position operator}position operators for all the nuclei}. This statement is true in any representation of the operators, and in particular must be respected if one chooses a position representation.

This is \textit{not} what \index{Born}Born and \index{Oppenheimer}Oppenheimer assumed about their equation (12)  [our equation(\ref{Scheq})] when $\kappa~=~0$ - see \S \ref{BOtheory} above - and which has been assumed ever since in Quantum Chemistry. In effect they chose to work only in the `small' \index{Hilbert space}Hilbert space of a fixed configuration, ${\cal{H}({\bf X})}$, in which ${\bf X}$ can be assumed to be a `parameter' in the position space wavefunction $\psi({\bf x},{\bf X})$, whereas if they had continued with quantum mechanics they would have been working in the `big' \index{Hilbert space}Hilbert space ${\cal{H}}$ with $\hat{\bf x}$ and $\hat{\bf X}$ treated on an equal footing as operators, and all possible nuclear configurations being treated simultaneously, rather than one at a time.

The unusual properties of the (`electronic') \index{Hamiltonian}Hamiltonian $\mathsf{\hat{H}}_o(\hat{\bf x},\hat{\bf p},\hat{\bf X})~=~\mathsf{\hat{H}}^{\mbox{elec}}$ in equation (\ref{decomp})\footnote{We assume that the centre-of-mass contributions are eliminated as usual.} considered as a quantum-mechanical operator on the whole space $\mathfrak{H}$, are of exactly the kind to be expected from the work of Kato \cite{K:51}. In Lemma 4 of his paper he showed that for a \index{Coulomb}Coulomb potential ${U}$ and for any function $f$ in the domain $\mathcal{D}_0$ of the full \index{kinetic energy}kinetic energy operator $\mathsf{\hat{T}}_0$, the domain, $\mathcal{D}_U$, of the internal \index{Hamiltonian}Hamiltonian $\mathsf{\hat{H}}'$ contains $\mathcal{D}_0$ and there are two constants $a,~b$ such that 
\[ ||{U}f|| \leq a||\mathsf{\hat{T}}_0 f|| +b||f|| \]
where $a$ can be taken as small as is liked.  This result is often
summarised by saying that the \index{Coulomb}Coulomb potential is small compared to
the \index{kinetic energy}kinetic energy. Given this result he proved in Lemma 5 (the
Kato-Rellich theorem) that the usual \index{Coulomb}Coulomb \index{Hamiltonian}Hamiltonian operator is essentially \index{self-adjoint}self-adjoint and so is guaranteed a complete set of eigenfunctions, and is bounded from below.  

In the present context the important point to note is that the \index{Coulomb}Coulomb
term is small only in comparison with the \index{kinetic energy}kinetic energy term
involving the same set of variables. So the absence of one or more 
\index{kinetic energy}kinetic energy terms from the
\index{Hamiltonian}Hamiltonian may mean that the
\index{Coulomb}Coulomb potential term cannot be treated as small. It
is evident that one can't use the Kato-Rellich argument to guarantee
\index{self-adjoint}self-adjointness for the customary representation
of $\mathsf{H}^{\mbox{elec}}$ in a position representation as a
differential and multiplicative operator because it contains the
nuclear positions  \{${\bf X}$\} in
\index{Coulomb}Coulomb terms that are not dominated by corresponding
\index{kinetic energy}kinetic energy operators involving the conjugate
\index{momentum operator}momentum operators \{$-i\hbar{\boldsymbol{\nabla}}$\} since they
have been separated off into the `perturbation' term $\propto
\kappa^4$. 
As a quite separate matter, the abstract direct integral
operator (\ref{decomp}) \textit{is} self-adjoint since the resolvent
of the clamped-nuclei Hamiltonian is integrable. This is demonstrated
in Theorem XIII.85 in the book by Reed and Simon \cite{RS}. It is in
this form that the operator is used in the mathematically rigorous
accounts (to be discussed later) of the Born-Oppenheimer
approximation in \cite{CDS:81} and \cite{KMSW:92}. The operator used
in the standard account of Born and Huang \cite{BHu} is however simply
the usual one which, as discussed above, is not self-adjoint in the
Kato sense.

\subsection{Approximate calculations}
\label{approx}
It might have been hoped, in the light of the claim in the original
paper by \index{Born}Born and \index{Oppenheimer}Oppenheimer quoted in \S \ref{BOtheory}, that the
eigensolutions of the $\kappa~\rightarrow~0$ limit of the internal \index{Hamiltonian}Hamiltonian, $\mathsf{\hat{H}}'$, would actually be those that would have
been obtained from (\ref{Scheq}) after separation of the centre-of-mass term, by letting the nuclear masses increase without limit. Although there are no analytically solved molecular problems, the work of Frolov 
\cite{FROL:99} provides extremely accurate numerical solutions for a
problem with two nuclei and a single electron. Frolov investigated
what happens when the masses of one and then two of the nuclei
increase without limit in his calculations. To appreciate his results,
consider a system with two nuclei; the natural nuclear coordinate is
the internuclear distance which will be denoted here simply as
$\mat{t}$.  When needed to express the electron-nuclei attraction
terms, $\mat{x}^{\rm n}_i$ is simply of the form  
$\alpha_i\mat{t}$ where $\alpha_i$ is a signed ratio of the nuclear mass
to the total nuclear mass; in the case of a homonuclear system
$\alpha_i=\pm\frac{1}{2}$. 

The di-nuclear electronic \index{Hamiltonian}Hamiltonian after the elimination of the centre-of-mass contribution as described in \S \ref{FQT} is 
\begin{eqnarray}
\!\!\!\!\mathsf{\hat{H}}^{\mbox{elec}}({\mat t}^{\rm e},{\mat{t}}) &=& -\frac{\hbar^2}{2m}\sum_{i=1}^{N} 
{\nabla}^2({\mat t}^{\rm e}_i) -\frac{\hbar^2}{2(m_1+m_2)}\sum_{i, j=1}^{N} 
\vec{\nabla}(\mat{t}^{\rm e}_i)\cdot\vec{\nabla}(\mat{t}^{\rm e}_j)\nonumber\\&-&\frac{e^2}{4\pi{\epsilon}_0}\sum_{j=1}^N \left(\frac{Z_1}{|{\mat t}^{\rm
e}_j+\alpha_1\mat{t}|}+\frac{Z_2}{|{\mat t}^{\rm e}_j+\alpha_2\mat{t}|}\right)\nonumber\\ &+&\frac{e^2}{8\pi{\epsilon}_0}\sum_{i,j=1}^N\!\hbox{\raisebox{5pt}{${}^\prime$}}\frac{1}{|{\mat t}^{\rm e}_i - {\mat t}^{\rm e}_j|}+\frac{Z_1Z_2}{R},~~R=|\mat{t}| 
\label{dineh}
\end{eqnarray}
while the nuclear \index{kinetic energy}kinetic energy part is:
\be
\mathsf{\hat{T}}_{\mbox{N}}(\mat t)~=~-\frac{\hbar^2}{2}\left(\frac{1}{m_1}+\frac{1}{m_2}\right){\nabla}^2({\mat t}) \equiv-\frac{\hbar^2}{2\mu}{\nabla}^2({\mat t}). 
\label{dinke}
\ee
The full internal motion \index{Hamiltonian}Hamiltonian for the three-particle system is then
\be
\mathsf{\hat{H}}'({\mat t}^{\rm e}, \mat{t})~=~\mathsf{\hat{H}}^{\mbox{elec}}({\mat t}^{\rm e},{\mat{t}}) ~+~\mathsf{\hat{T}}_{\mbox{N}}(\mat{t}). 
\label{intH}
\ee
which is of the same form as (\ref{intHam}).

It is seen from (\ref{dinke}), that if only one nuclear mass increases
without limit then the \index{kinetic energy}kinetic energy term in the nuclear variable
remains in the full problem and so the \index{Hamiltonian}Hamiltonian (\ref{intH}) remains
essentially \index{self-adjoint}self-adjoint. Frolov's calculations showed that when one mass increased without limit (the atomic case), any discrete
spectrum persisted but when two masses were allowed to increase
without limit (the molecular case), the \index{Hamiltonian}Hamiltonian ceased to be
well-defined and this failure led to what he called \textit{\index{adiabatic}adiabatic
divergence} in attempts  to compute discrete eigenstates of
(\ref{intH}). This divergence is discussed in some mathematical detail in the Appendix to Frolov \cite{FROL:99}. It does not arise from the choice of a translationally invariant form for the electronic \index{Hamiltonian}Hamiltonian; rather it is due to the lack of any \index{kinetic energy}kinetic energy term to dominate the \index{Coulomb}Coulomb potential. 

To every solution of (\ref{HoSch}) there corresponds a function
\begin{equation}
\Phi({\bf t}^{\rm e},{\bf t}^{\rm n})_m~=~\varphi({\bf b},{\bf t}^{\rm
e})_m~\delta({\bf t}^{\rm n}~-~{\bf b})
\label{nonorm}
\end{equation}
in the $({\bf t}^{\rm e},{\bf t}^{\rm n})$ position representation which is a formal solution, in the sense of distributions, of the \index{Schr\"{o}dinger}Schr\"{o}dinger equation for $\mathsf{\hat{H}}^{\mbox{elec}}$.  The energy, ${\cal{E}}_m({\bf b})$ of the function (\ref{nonorm}) is independent of the orientation of the figure defined by the ${\bf b}$, and is also unaltered by the parity operation ${\bf b}~\rightarrow~-{\bf b}$, and by permutations of the labelling of any identical nuclei. $\Phi_m$ however depends on the orientation of the body-fixed frame defined by the configuration ${\bf b}$ with respect to some space-fixed reference frame. Let the Euler angles relating these two frames be $\Omega$ so that
\begin{displaymath}
\Phi({\bf b})_m~=~\Phi({\overline{\bf b}},\Omega)_m
\end{displaymath}
in an obvious notation, so we have a continuous family of degenerate states. The dependence on orientation is eliminated by forming a continuous superposition through integration over the Euler angles with some weight function $c(\Omega)$
\begin{displaymath}
\Psi_m~=~\int d\Omega'~c(\Omega')~\Phi({\overline{\bf b}},\Omega')_m
\end{displaymath}
Similarly one may form superpositions of the space-inverted and permuted states in order to form a new basis that displays the corresponding symmetries that leave the energy eigenvalue unchanged.

There are two quite distinct approaches to the solution of the
molecular \index{Schr\"{o}dinger}Schr\"{o}dinger equation (\ref{intScheq}) based on the formal theory reviewed in \S \ref{FQT}. Functions of the type (\ref{nonorm}) can be used as the basis of a Rayleigh-Ritz calculation being, hopefully, well-adapted to the construction of useful trial functions. Several different lines have been developed; in the \textit{\index{adiabatic}adiabatic} model the trial function is written as the continuous linear
superposition
\begin{eqnarray}
 \Psi({\bf t}^{\rm e},{\bf t}^{\rm n})_m~&=&~\int
d{\bf b}~F({\bf b})~\varphi({\bf b},{\bf t}^{\rm e})_m~\delta({\bf
t}^{\rm n}~-~{\bf b})\nonumber\\
&=&F({\bf t}^{\rm n})~\varphi({\bf t}^{\rm n},{\bf t}^{\rm e})_m
\label{trialpsi}
\end{eqnarray}
where the square-integrable weight factor $F({\bf t}^{\rm n})$ may
be determined by reducing (\ref{intScheq}) to an effective
\index{Schr\"{o}dinger}Schr\"{o}dinger equation for the nuclei in which $F({\bf t}^{\rm
n})$ appears as the eigenfunction \cite{am60}. 

If the \{$\varphi_m$\} are chosen to be orthonormal we have
\begin{displaymath}
\langle \Psi_m|\Psi_m\rangle~=~\int\int d{\bf t}^{\rm e}~d{\bf
t}^{\rm n}~|\Psi({\bf t}^{\rm e},{\bf t}^{\rm n})_m|^2~=~\int d{\bf
t}^{\rm n}~|F({\bf t}^{\rm n})|^2
\end{displaymath}
We may choose the weight factor $F$ to be normalized, so that the state function $\Psi_m$ is also normalized. On the other hand
\begin{equation}
\begin{aligned}
\langle\Psi_m|\mathsf{\hat{H}}'|\Psi_m\rangle~&=~\int\int d{\bf t}^{\rm
e}~d{\bf t}^{\rm n}~\Psi_{m}^{*}\left(\mathsf{\hat{H}}'~\Psi_m\right)\\~&=~\int
d{\bf t}^{\rm n}~F({\bf t}^{\rm n})^*\left(\mathsf{\hat{H}}_m~F\right)({\bf
t}^{\rm n})
\label{nucleq}
\end{aligned}
\end{equation}
where we have defined the effective nuclear \index{Hamiltonian}Hamiltonian
\begin{equation}
\left(\mathsf{\hat{H}}_m~F\right)({\bf t}^{\rm n})~=~\int d{\bf t}^{\rm
e}~\varphi({\bf t}^{\rm e},{\bf t}^{\rm
n})_m\left[\mathsf{\hat{H}}'~\varphi({\bf t}^{\rm e},{\bf t}^{\rm
n})_m~F({\bf t}^{\rm n})\right]
\label{nucleqn}
\end{equation}
The Rayleigh-Ritz quotient
\begin{equation}
\label{RayR}
E[\Psi_m]~=~\frac{\langle\Psi_m|\mathsf{\hat{H}}'|\Psi_m\rangle}{\langle
\Psi_m|\Psi_m\rangle}
\end{equation}
is stationary for those functions that are solutions of the
effective nuclear `\index{Schr\"{o}dinger}Schr\"{o}dinger equation'
\begin{equation}
\mathsf{\hat{H}}_m~F_s~=~E_{ms}~F_s
\label{effnucl}
\end{equation}

In particular, using the electronic ground state $\varphi_0$, the
Rayleigh-Ritz quotient leads to an \textit{upper bound} to the ground
state energy $E_0$ of $\mathsf{\hat{H}}'$. Having set up the calculation with square integrable functions the approximate ground-state is naturally a discrete state; the discussion however yields no information about the bottom of the essential spectrum i.e. it does not prove the existence of a bound-state below the continuum. This calculation amounts to the diagonalization of the projection of $\mathsf{\hat{H}}'$ on the one-dimensional subspace spanned by $\Psi_0$. In principle the subspace may be enlarged, and the accuracy thereby improved, by using the subspace spanned by a set of trial functions ($\Psi_0,\Psi_1,\cdots,\Psi_m$) of the form of (\ref{trialpsi}). Such non-\index{adiabatic}adiabatic calculations which make no use of a \index{Potential Energy Surface}Potential Energy Surface are restricted to very small molecules.  

In practice the variational approach is implemented as follows; a collection of energies $E({\bf b}_i)$ is found through standard quantum chemical computations for different geometries \{${\bf b}_i$\} and fitted to produce a function $V({\bf t}^{\rm n})$ that is treated as a potential energy contribution to the left-hand-side of the \index{Born}Born equation (\ref{coupl}), rather than (\ref{effnucl}), so the \index{clamped-nuclei} clamped-nuclei assumption enters in an essential way (see Appendix). With considerable computational effort it is possible to construct permutationally invariant energy surfaces for molecules with up to 10 nuclei \cite{BB:09}. Note however that if $\mathsf{\hat{H}}'$ is separated as in (\ref{intHam}), then it is $\mathsf{\hat{H}}^{\mbox{elec}}$ that appears in (\ref{nucleqn}) rather than the \index{clamped-nuclei} clamped-nuclei \index{Hamiltonian}Hamiltonian.

Another generalization is to replace the unnormalizable delta function in (\ref{trialpsi}) by a square integrable function; the relation
\begin{displaymath}
\delta^3({\bf x}~-~{\bf
y})~=~\lim_{a\rightarrow\infty}\left(\frac{a}{\pi}\right)^{\frac{3}{2}}~e^{-a({\bf
x}~-~{\bf y})^2}~\equiv~\lim_{a\rightarrow\infty}\chi_a({\bf x},{\bf
y})
\end{displaymath}
suggests that one might consider trial wavefunctions
\begin{displaymath}
\Psi({\bf t}^{\rm e},{\bf t}^{\rm
n})_{m}^{\mbox{\scriptsize{GCM}}}~=~\int d{\bf b}~F({\bf
b})~\varphi({\bf b},{\bf t}^{\rm e})_m~\chi_a({\bf t}^{\rm n},{\bf
b})
\end{displaymath}
for some suitably chosen parameter $a$. This is the basis of the
molecular \textit{\index{Generator Coordinate Method}Generator Coordinate Method} (GCM) which is a
non-\index{adiabatic}adiabatic formalism; as before the weight factor $F({\bf b})$ is
determined by appeal to the Rayleigh-Ritz quotient, although part of its structure can be determined purely by symmetry arguments. In the
GCM the effective \index{Schr\"{o}dinger}Schr\"{o}dinger equation for the weight function
becomes an integral equation (the Hill-Wheeler equation)
\cite{LVL}. Again the trial function may be improved, in the sense
of a variational calculation, by forming linear superpositions of
the wavefunctions \{$\Psi^{\mbox{\scriptsize{GCM}}}$\}; this has been
done for diatomic molecules for which a fairly complete GCM account
has been developed \cite{LVL},\cite{BLvL}. Usually however the dependence on the nuclear variables \{${\bf t}^{\rm n}$\} is not expressed through
functions adapted to nuclear permutation symmetry, and the GCM weight functions are determined by \index{molecular structure}molecular structure considerations.

It should be noted here that $\varphi({\bf b},{\bf t}^{\rm e})$ as a
solution to the \index{Schr\"{o}dinger}Schr\"{o}dinger equation (\ref{HoSch}) where ${\bf
t}^{\rm n}$ has been replaced by ${\bf b}$, is defined only up to a phase
factor of the form \[\exp[iw({\bf b})] \] $w$ is any single-valued real function of the \{${\bf b}_i$\} which can be different for different electronic states. The phase factor is only trivial in the absence of degeneracies. Specific phase choices may therefore be needed when tying this part to the nuclear part of the product wave function. It is only by making suitable phase choices that the electronic wave function is made a
continuous function of the formal nuclear variables, ${\bf b}$, and
the complete product function, made single valued. This is the
origin of the Berry phase in \index{clamped-nuclei} clamped-nuclei calculations involving
intersecting \index{Potential Energy Surface}Potential Energy Surfaces; for a discussion of these
matters see \cite{CAM:92}, \cite{FZ:02}. It is worth noting explicitly that
notions of molecular Berry phases and conical intersections of PE
surfaces are tied to the \index{clamped-nuclei} clamped-nuclei viewpoint which introduces `\index{adiabatic}adiabatic parameters'. According to quantum mechanics the
eigensolutions of (\ref{intScheq}) are single-valued functions by
construction with arbitrary phases (rays) so one does not expect any
Berry phase phenomena \textit{a priori}.

The rigorous mathematical analysis of the original perturbation
approach proposed by \index{Born}Born and \index{Oppenheimer}Oppenheimer \cite{BO} for a molecular
\index{Hamiltonian}Hamiltonian with \index{Coulomb}Coulombic interactions was initiated by Combes and
co-workers \cite{C:75} - \cite{CDS:81} with results for the diatomic
molecule. Some properties of the operator $\mathsf{H}^{\mbox{elec}}$, (equation \ref{decomp}), seem to have been first discussed in this work. A perturbation expansion in powers of $\kappa$ leads to a
singular perturbation problem because $\kappa$ is a coefficient of
differential operators of the highest order in the problem; the
resulting series expansion of the energy is an \textit{asymptotic}
series, closely related to the WKB approximation obtained by a  semiclassical analysis of the effective \index{Hamiltonian}Hamiltonian for the nuclear dynamics. This requires a more complete treatment than the \index{adiabatic}adiabatic model using the partitioning technique to project the full \index{Coulomb}Coulomb \index{Hamiltonian}Hamiltonian, $\mathsf{\hat{H}}'$, onto the \index{adiabatic}adiabatic subspace. A normalized electronic eigenvector
$|\varphi({\bf b})_j\rangle$ is associated with a projection
operator by the usual correspondence
\begin{displaymath}
\hat{P}({\bf b})_j~=~|\varphi({\bf b})_j\rangle \langle\varphi({\bf
b})_j|
\end{displaymath}
In view of our earlier discussion of the `big \index{Hilbert space}Hilbert space'
${\cal{H}}$, we can form a \index{direct integral}direct integral over all nuclear
positions
\begin{displaymath}
\hat{P}_j~=~\int^{\oplus}_{X} ~\hat{P}({\bf b})_j~d{\bf b}
\end{displaymath}
to yield a projection operator on the \index{adiabatic}adiabatic subspace. If we want
to include $m$ electronic levels we can form a direct sum of the
contributing \{$\hat{P}_j$\}
\begin{displaymath}
\hat{P}~=~\bigoplus_{j=0}^{m}~\hat{P}_j
\end{displaymath}
This is an Hermitian projection operator and it, and its complement,
$\hat{Q}$, have the usual properties
\begin{displaymath}
\hat{P}~+~\hat{Q}~=~\hat{1},~~~\hat{P}^2~=~\hat{P},~~~\hat{Q}^2~=~\hat{Q},~~~\hat{P}\hat{Q}~=~\hat{Q}\hat{P}~=~0
\end{displaymath}

Using these projection operators the original molecular
\index{Schr\"{o}dinger}Schr\"{o}dinger equation
\begin{displaymath}
\mathsf{\hat{H}}'|\Psi\rangle~=~E|\Psi\rangle
\end{displaymath}
can be transformed into a pair of coupled equations
\begin{eqnarray}
\label{projP}
\hat{P}\mathsf{\hat{H}}'\hat{P}|\psi\rangle~+~\hat{P}\mathsf{\hat{H}}'\hat{Q}|\chi\rangle~&=&~E\hat{1}|\psi\rangle\\
\label{projQ}
\hat{Q}\mathsf{\hat{H}}'\hat{P}|\psi\rangle~+~\hat{Q}\mathsf{\hat{H}}'\hat{Q}|\chi\rangle~&=&~E\hat{1}|\chi\rangle
\end{eqnarray}
where
\begin{displaymath}
|\psi\rangle~=~\hat{P}|\Psi\rangle,~~~|\chi\rangle~=~\hat{Q}|\Psi\rangle
\end{displaymath}
Solving (\ref{projQ}) for $|\chi\rangle$
\begin{displaymath}
|\chi\rangle~=~\frac{1}{E\hat{1}~-~\hat{Q}\mathsf{\hat{H}}'\hat{Q}}~\hat{Q}\mathsf{\hat{H}}'\hat{P}|\psi\rangle
\end{displaymath}
and substituting in (\ref{projP}) yields the usual \index{L\"{o}wdin}L\"{o}wdin partitioned equation \cite{POL:66}
\begin{equation}
\left(\hat{P}\mathsf{\hat{H}}'\hat{P}~+~\hat{P}\mathsf{\hat{H}}'\hat{Q}\frac{1}{E\hat{1}~-~\hat{Q}\mathsf{\hat{H}}'\hat{Q}}\hat{Q}\mathsf{\hat{H}}'\hat{P}\right)|\psi\rangle~=~E\hat{1}|\psi\rangle
\label{Loweqn}
\end{equation}

Further progress depends crucially on establishing the properties of
the energy dependent operator in (\ref{Loweqn}). A detailed
consideration of the diatomic molecule case can be found in
\cite{CS:80}, \cite{CDS:81}. The main result is that (\ref{Loweqn}) is a
generalized version of the effective nuclear \index{Schr\"{o}dinger}Schr\"{o}dinger
equation (\ref{effnucl}) in the \index{adiabatic}adiabatic model, so it contains the
nuclear \index{kinetic energy}kinetic energy operators and an effective potential
$\hat{V}$. The \textit{ \index{microscope transformation}microscope transformation} used by Combes and
Seiler \cite{CS:80} to give a rigorous version of the \index{Born}Born-\index{Oppenheimer}Oppenheimer
theory of a diatomic molecule is essentially a semiclassical theory.
It is applicable if there is a minimum in the potential
$V_{\mbox{0}} = V({\bf x}_ 0)$ associated with a particular
configuration of the nuclei\footnote{ The multiminima case can also
be treated in this way.} that is deep enough for the lowest energy
eigenstates to be localized about ${\bf x}_0$. One can look at these
states with a `microscope' with a certain resolving power that
depends on Planck's constant.

The \index{microscope transformation}microscope transformation produces a translation to make ${\bf x}_0$ the origin of the coordinates, and a dilation (scale transformation)
\begin{equation}
\hat{\bf x}_{\lambda}~=~\hat{\bf x}~-~(1-\lambda)\, (\hat{\bf
x}~-~{\bf x}_0),~~~\hat{\bf p}_{\lambda}~=~\hat{\bf p}~+~~\frac{(1-\lambda)}{\lambda} \hat{\bf p} \label{lamderiv}
\end{equation}
It is readily verified that the commutation relations are preserved for $\lambda~\neq~0$
\begin{displaymath}
[\hat{\bf x}_{\lambda},\hat{\bf p}_{\lambda}]~=~[\hat{\bf x},\hat{\bf p}]
\end{displaymath}
Under this transformation a \index{Hamiltonian}Hamiltonian of the form
\begin{displaymath}
\hat{H} ~=~\sum_g \frac{\hat{p}_{g}^2}{2 m_g}~+~ \hat{V} ({\bf x})
\end{displaymath}
becomes
\begin{displaymath}
\hat{H}_{\lambda} ~=~\hat{V}({\bf x}_0)~+~\lambda^{2}
\hat{N}(\lambda )
\end{displaymath}
where
\begin{displaymath}
\hat{N}(\lambda)~=~-~\frac{\hbar^2}{\lambda^{4}}\sum_g
\frac{\nabla_{g}^{2}}{2 m_g}~+~\frac{1}{\lambda^2}\,\, \left(\hat{V}
\big({\bf x}_0~+~\lambda({\bf x}~-~{\bf x}_0)\big)~-~\hat{V}({\bf x}_0)
\right)
\end{displaymath}

We now put $\lambda$ = $\surd\hbar$ so as to eliminate $\lambda$
from the \index{kinetic energy}kinetic energy term in $\hat{N}(\lambda)$; with this choice
for $\lambda$, unitary equivalence of the spectrum implies that the
eigenvalues of the original \index{Hamiltonian}Hamiltonian $\hat{H}$ are related to
those of $\hat{N}(\lambda)$ by
\begin{displaymath}
E_n ~=~V{(\bf x_0)} ~+~ \hbar \mu_n(\lambda)
\end{displaymath}
Provided $\hat{V}$ is analytic in $\lambda$ it can be expanded about
$\lambda = 0$, and this puts $\hat{N}(\lambda)$, in lowest order,
into the form of a sum of coupled oscillators so that the first
approximation for the eigenvalue function $\mu_n$ is
\begin{displaymath}
\mu_n~=~\sum_k\big(n_k~+~\tfrac{1}{2}\big)
\end{displaymath}
In the \index{Born}Born-\index{Oppenheimer}Oppenheimer calculation for the diatomic molecule the
potential $\hat{V}$ is identified with the effective potential for
the nuclei \cite{CS:80}; analyticity of $\hat{V}$ in $\lambda$ could be
proven, and the role of $\surd\hbar$ was taken by the usual BO
expansion parameter $\kappa = (m_e/M_N)^{\frac{1}{4}}$. In this way
the molecular energy level formula (\ref{Emol}) is recovered as
an asymptotic expansion.

The singular nature of the \index{microscope transformation}microscope transformation for $\lambda~=~0$ is demonstrated by the modification of the spectrum associated with the limit
$\kappa \rightarrow 0$. The spectrum of the \index{Coulomb}Coulomb \index{Hamiltonian}Hamiltonian for
a molecule can be discussed in terms of the Hunziker-van Winter-Zhislin theorem \cite{GZ:60}-\cite{WH:66}; for the diatomic molecule, $\sigma_{ess}(\mathsf{\hat{H}}')$ starts at the lowest two-body threshold $\Sigma = \lambda_A(m_A) + \lambda_B(m_B)$ given by the minimal value of the sums of pairs of binding energies for atoms A and B with finite masses $m_A$ and $m_B$ respectively. On the other hand the spectrum of the electronic \index{Hamiltonian}Hamiltonian,
$\mathsf{\hat{H}}^{\mbox{elec}}$, is purely continuous, $\sigma(\mathsf{\hat{H}}^{\mbox{elec}}) = [{V}_0, \infty)$. In the limit $m_A, m_B\rightarrow \infty$, $\Sigma$ does not generally converge to ${V}_0$; instead the
missing part of the \index{continuous spectrum}continuous spectrum [${V}_0$,
$\lambda_A(\infty) + \lambda_B(\infty)$] is provided by an
accumulation of bound states in this interval \cite{C:77}. The \index{microscope transformation}microscope transformation is formally applicable to the polyatomic case but it may not be sufficient to control the asymptotic behaviour, and has not been used for general molecules.

Since the initial work of Combes, a considerable amount of mathematical work has been published using both time-independent and time-dependent techniques with developments for the polyatomic case; for a recent review of rigorous results about the separation of electronic and nuclear motions see Hagedorn and Joye \cite{HJ:07} which covers the literature to 2006.  The \index{Hamiltonian}Hamiltonian (\ref{intHam}) is the one used by Klein
\textit{et al.} \cite{KMSW:92} in their consideration of the precise
formulation of the \index{Born}Born-\index{Oppenheimer}Oppenheimer approximation for polyatomic
systems. Their work was based on a powerful symbolic operator
method, the \textit{pseudodifferential calculus} \cite{KMSW:92}, \cite{CLF:83} and a formalism related to the partitioning technique described above. In \cite{KMSW:92} it is assumed that (\ref{intHam}) has a discrete
eigenvalue which has a minimum as a function of the ${\bf t}^{\rm
n}$ in the neighborhood of some values ${\bf t}^{\rm n}_i={\bf
b}_i$. If it can be assumed that a) the electronic wavefunction
vanishes strongly outside a region close to a particular nuclear
geometry and b) that the electronic energy at the given geometry is an
isolated minimum, then it is possible to present a rigorous account of the separation of electronic and nuclear motion which corresponds in some measure to the original \index{Born}Born-\index{Oppenheimer}Oppenheimer treatment. 

A novel feature arises from the requirement that the inversion symmetry of the original problem be respected. If the geometry at the minimum energy configuration is either linear or planar then inversion can be dealt with in terms of a single minimum in the electronic energy. If the geometry at the minimum is other than these two forms, inversion produces a second potential minimum and the problem must be dealt with as a two-minimum problem; then extra consideration is necessary to establish whether the two wells have negligible interaction so that only one of the wells need be
considered for the nuclear motion. The nuclei are treated as
distinguishable particles that can be numbered uniquely. The symmetry
requirements on the total wavefunction that would arise from the
invariance of the \index{Hamiltonian}Hamiltonian operator under the permutation of
identical nuclei are not considered. Because of the \index{continuous spectrum}continuous spectrum of the electronic \index{Hamiltonian}Hamiltonian $\mathsf{\hat{H}}^{\mbox{elec}}$, it is not possible to use regular perturbation theory in the analysis; instead asymptotic expansion theory is used so that the result has essentially the character of a WKB approximation \cite{KMSW:92}. Similar functional analytic techniques have been used to consider such phenomena as Landau-Zener crossing by using a time-dependent approach to the problem and looking at the relations between the electronic and nuclear parts of a wave-packet \cite{HJ:99}. This is essentially a use of standard coherent state theory where again the nuclei are treated as distinguishable particles and the method is that of asymptotic expansion.

\section{Discussion}
\label{Disc}

Quite generally one needs to make a distinction between an hypothetical \index{Isolated Molecule}Isolated Molecule, and a real observable \textit{individual} molecule. There are no strictly `isolated' systems of course, but what is striking is that an approach based on the \index{stationary state}stationary state eigensolutions of the appropriate \index{Coulomb}Coulomb \index{Hamiltonian}Hamiltonian works so well for atoms and diatomic molecules, but fails with three or more nuclei. We have always been clear that for most of chemistry, molecular eigenstates (`\index{stationary state}stationary states') are of no relevance since \textit{metastability} is an essential aspect of isomerism. The interesting question is how to get from the quantum theory of an \textit{\index{Isolated Molecule}Isolated Molecule} to a quantum theory of an \textit{individual molecule} by rational mathematics. It is as well to remember that the \textit{generic} molecule is sufficiently complex that the quantum mechanical permutation symmetry of identical nuclei is a feature that cannot be ignored, if one is doing quantum mechanics. The \index{Isolated Molecule}Isolated Molecule model doesn't capture isomerism, nor optical activity. We see no reason at all for \index{L\"{o}wdin}L\"{o}wdin's optimistic assertion (\S \ref{Intro}) that molecular symmetry \textit{must} be contained somehow in the \index{Coulomb}Coulomb \index{Hamiltonian}Hamiltonian. 

If a molecule is not isolated it must be interacting with something; that something is loosely referred to as the `environment'. It might be other molecules, the (macroscopic) substance the molecule finds itself in, or quantized electromagnetic radiation. Blackbody radiation is all pervasive and charges are always coupled to the photon vacuum state in QED and so `dressed' with clouds of virtual photons. A crucial feature of `environments' in quantum theory is that generally they are described by \index{Hamiltonian}Hamiltonians with purely continuous spectra. This is important because a quantum system with a finite number, $n$, of degrees of freedom described by the usual linear \index{Schr\"{o}dinger}Schr\"{o}dinger equation does not yield `\index{broken symmetry}broken symmetry' solutions if $n~<~\infty$. Such matters were discussed at length thirty years ago in the context of \index{molecular structure}molecular structure and quantum theory \cite{HP:83} - \cite{RGW:82}. The characteristic feature of such discussions, and this also applies to more modern formulations under the chic heading of `decoherence', is that they start with some primitive notion of structure built in: two-state systems, potential energy wells, wavefunctions associated with distinct isomers \textit{etc.}. The `environment' is modelled in the simplest possible way (for example, a free boson quantum field). These crucial ideas are put in by hand at the outset. We don't see that as a `problem' or `difficulty'; it is a characteristic feature of many-body physics (condensed matter, nuclei, chemistry) and results in remarkably powerful and fruitful theoretical formalisms; see, for example, Anderson's discussion of what he calls `\index{adiabatic}adiabatic continuity' \cite{PWA:84}. But one can hardly avoid noticing that the models of molecules used are caricatures that contain just the right features for the answer (for example, loss of permutation symmetry, loss of parity - in the case of chirality) and have no clear connection to \index{L\"{o}wdin}L\"{o}wdin's's \index{Coulomb}Coulomb \index{Hamiltonian}Hamiltonian.

An alternative account that \textit{is} based on the \index{Coulomb}Coulomb \index{Hamiltonian}Hamiltonian may however be possible in the light of the fact that each part of it in the division made in (\ref{intHam}) has a completely \index{continuous spectrum}continuous spectrum. As noted in \S \ref{approx} the formal eigenvectors of $\mathsf{\hat{H}}^{\mbox{elec}}$ from the ground state up can exhibit extensive degeneracy. It might be that `\index{broken symmetry}broken symmetry' solutions corresponding to \index{molecular structure}molecular structure could result from treating the two parts as asymptotic states in a scattering or reaction process in a manner analogous to that used in standard S-matrix theory. Such a state would be characterised as a `\index{resonance}resonance' and would have to be long-lived to be describable as a molecule. Only a true bound state, of infinite lifetime, such as a bound  eigenstate of the molecular \index{Hamiltonian}Hamiltonian is really independent of how it was formed; resonant states have histories that describe the environment of their preparation. The \index{Potential Energy Surface}Potential Energy Surface would then appear only as an auxiliary concept through the involvement of the \index{clamped-nuclei} clamped-nuclei \index{Hamiltonian}Hamiltonian in the construction of the states \{$\varphi_{m}$\} required for equation (\ref{nonorm}).  That however remains subject to further investigation.

When \index{Bohr}Bohr introduced his quantum theory of the Rutherford atomic model of the hydrogen atom he made a drastic change in the status of \index{electrodynamics}electrodynamics. Hitherto, it had been understood\footnote{The Lorentz Theory of the electron for example \cite{HAL:16}.} that charged particles affected, and were affected by, the electromagnetic field, and that was the root cause of the failure of a dynamical classical atom (`radiation damping' is a strong coupling interaction). \index{Bohr}Bohr relegated the electromagnetic field to a perturbation theoretic - \textit{weak coupling} - role; the charges interacted among themselves according to \index{Coulomb}Coulomb's law, to be treated as a strong coupling situation, and would exist permanently in the \index{stationary state}stationary states selected by the quantization conditions unless perturbed by an `external' electromagnetic field which produced `transitions'. That perturbation theory viewpoint was maintained when quantum theory was applied to the atom and the electromagnetic field, and largely survives to this day, to the extent that the electromagnetic field is frequently regarded as a \textit{classical} system. Such a spectroscopic viewpoint is not appropriate in the present context; quantum \index{electrodynamics}electrodynamics teaches us that there is no strict separation of charged particles and the (quantized) electromagnetic field, not least because of the requirements of \index{gauge invariance}gauge invariance.

The difficulty with quantum mechanical perturbation theory for the interaction of atomic/molecular systems with radiation is this: the spectrum of the unperturbed atom/molecule consists of a continuum corresponding to the half-axis [$\Sigma,\infty$) for some $\Sigma ~\leq~0$, and discrete energy levels $E_0,E_1,\ldots$ below the continuum, that is $E_0\leq E_1 \leq,\ldots,<~\Sigma$ \cite{GZ:60} - \cite{WH:66}. The spectrum of the free electromagnetic field \index{Hamiltonian}Hamiltonian consists of a simple eigenvalue at $0$, corresponding to the vacuum state $\Psi_0$, and absolutely \index{continuous spectrum}continuous spectrum on the half-axis [$0,\infty$). This means that when coupling between particles and radiation is admitted, all the discrete energy levels of the atomic system including $E_0$ become thresholds of continuous spectra; a quantum theory of matter and the electromagnetic field therefore requires the perturbation theory of continuous spectra. The quantum mechanical theory of electrons and nuclei interacting with quantized radiation in the low-energy regime is an active area of research in mathematical physics concerned with the stability of matter, the existence of the thermodynamic limit \textit{etc.}, but with no particular reference to features of chemical interest \cite{LL:03} - \cite{LS:10}.

The presentation of a presumed exact bound state solution of the
\index{Schr\"{o}dinger}Schr\"{o}dinger \index{Coulomb}Coulomb \index{Hamiltonian}Hamiltonian as a product of electronic and
nuclear factors has been considered both by Hunter \cite{H} and,
more recently, by Gross et al. \cite{gross}. For present purposes the Hunter
approach will be employed on the translationally invariant form of the internal \index{Hamiltonian}Hamiltonian, given earlier (in \S \ref{FQT}). Were the exact solution known, Hunter argues that it could be written in the form
\be
\psi({\mat t}^{\rm n}, {\mat t}^{\rm e})=\chi(\mat{t}^{\rn})\phi({\mat
  t}^{\rm n}, {\mat t}^{\rm e}) 
\label{hunxct}
\ee
with the nuclear wave function defined by means of
\be
|\chi(\mat{t}^{\rn})|^2=\int\psi({\mat t}^{\rm n}, {\mat t}^{\rm
  e})^*\psi({\mat t}^{\rm n}, {\mat t}^{\rm e})d\mat{t}^{\re} 
\label{hunf}
\ee
Providing the function $\chi(\mat{t}^{\rn})$ has no nodes,\footnote {A similar requirement must be placed on the denominator in equation (12) of \cite{vrkut} for the equation to provide a secure definition} an `exact' electronic wavefunction could be constructed as 
\be
\phi({\mat t}^{\rm n}, {\mat t}^{\rm e})=\frac{\psi({\mat t}^{\rm
    n}, {\mat t}^{\rm e})}{\chi(\mat{t}^{\rn})} 
\label{hunewf}
\ee
if the normalization choice
\[
\int\phi({\mat t}^{\rm n}, {\mat t}^{\rm
  e})^*\phi({\mat t}^{\rm n}, {\mat t}^{\rm e})d\mat{t}^{\re}=1
\]
is made. The electronic wavefunction (\ref{hunewf}) is then properly defined, and a `\index{Potential Energy Surface}Potential Energy Surface' could be defined in terms of it by integrating out the electronic variables in the expectation value of the internal \index{Hamiltonian}Hamiltonian in the state $\phi$,
\be
\mathsf{U}({\mat t}^{\rm n})= \int\phi({\mat t}^{\rm n}, {\mat t}^{\rm
  e})^*\mathsf{\hat{H}}'({\mat t}^{\rm n}, {\mat t}^{\rm
  e}) \phi({\mat t}^{\rm n}, {\mat t}^{\rm
  e})d\mat{t}^{\re}
\label{hunpes}
\ee
The nuclear function $\chi$ is evidently quite different \cite{H1} from the usual approximate nuclear wavefunctions for vibrationally excited states which do have nodes.

Although no closed solutions to the full problem are known for a
molecule, some extremely good approximate solutions have been obtained for
excited vibrational states of H$_2$; Czub and Wolniewicz \cite{CW}
took such an accurate approximation for an excited vibrational
state in the $J=0$ rotational state of H$_2$ and computed
$\mathsf{U}(R)$. They found strong spikes in the potential close to
two positions at which the usual vibrational wave function would
have nodes. To quote \cite{CW}
\begin{quote}
This destroys completely the concept of a single internuclear potential in diatomic molecules because it is not possible to introduce on the basis of non-\index{adiabatic}adiabatic potentials a single, approximate, mean potential that would describe well more than one vibrational level.

It is obvious that in the case of rotations the situation is even more
complex.
\end{quote}

\index{Wilson}{Wilson} suggested \cite{ebw79} that using the \index{clamped-nuclei} clamped-nuclei
\index{Hamiltonian}Hamiltonian instead of the full one in (\ref{hunpes}) to define the
potential might avoid the spikes but Hunter in \cite{H1} showed
why this was unlikely to be the case, and Cassam-Chenai \cite{pcc06}
repeated the work of Czub and Wolniewicz using a \index{clamped-nuclei} clamped-nuclei electronic \index{Hamiltonian}Hamiltonian and showed that exactly the same spiky behaviour occurred.

Another approach to this problem is in \cite{gross}; there is reason to believe however that this sort of difficulty is bound to arise whatever the approach. To see this, simply rewrite (\ref{hunxct}) to
recognise that the exact states will actually have definite quantum
numbers  according to their symmetry, so that it would be more
realistic to write
\be
\psi_{J M p r i}({\mat t}^{\rm n}, {\mat t}^{\rm
  e})=\chi_{J M P r i}(\mat{t}^{\rn})\phi_{J M p r i}({\mat 
  t}^{\rm n}, {\mat t}^{\rm e}) 
\label{hunxct1}
\ee
In the H$_2$ study cited the first four quantum numbers are of
no relevance, only $i$ remains and here $i$ labels the vibrational
states. There is thus every reason to expect that the best that can be
done from this approach is a distinct PES for each nuclear-motion state.

This anticipated  behaviour seems to be confirmed in very accurate calculations on H$_2$ \cite{pk09} for the electronic $\Sigma$
ground state of the molecule assumed to dissociate into two hydrogen
atoms in their ground states. That work shows that, for example, the
$J=0$ state supports just 14 vibrational states while the $J=15$ state
supports 10 and the $J=31$ supports only 1 state. Of course in a
diatomic molecule, states of different $k$ are states which differ in
the electronic angular momentum and these results cannot be regarded
as typifying the results for a polyatomic system. However work on
H$_3^+$ shows that in the case of $J=0$ there are 1280 vibrational
states below dissociation \cite{hts93} and that $46$ is the highest value of $J$ for which at least one vibrational state exists \cite{mt88}.
At this level then it cannot be assumed that the potential surface
calculated in the usual way is an approximation to anything exact.

The eigenstates of the full molecular \index{Hamiltonian}Hamiltonian (the \index{Coulomb}Coulomb \index{Hamiltonian}Hamiltonian for the electrons and nuclei specified by a chemical formula) - a theory of an \index{Isolated Molecule}Isolated Molecule modeled on the quantum theory of the atom which we call the \index{Isolated Molecule}Isolated Molecule model - are reasonably well understood and might have some utility in a limited area of high-resolution experiments on very small molecules where questions of isomerism do not arise \cite{WS:05}. Their computation poses formidable problems, and really belongs to few-body physics. If it is to be taken as underlying Quantum Chemistry then it is worth exploring the consequences of the model without regard to approximations made for practical utility which are a quite separate matter.  In this paper we have attempted to discuss the \index{Born}Born-\index{Oppenheimer}Oppenheimer and \index{Born}Born approaches to the quantum theory of molecules in terms first set out by Combes \cite{C:75}. The essential point is that the decomposition of the molecular \index{Hamiltonian}Hamiltonian (with centre-of-mass contribution removed) into the nuclear \index{kinetic energy}kinetic energy, proportional to $\kappa^4$ and a remainder, is specified by equation (\ref{intHam}), not by (\ref{fullH}), or in other words, \textit{equation (\ref{fullH}) cannot be written with an $=$ sign} if the conventional interpretation of the $X$ acting as parameters is made. Allowing the nuclear masses to increase without limit in
$\mathsf{\hat{H}}^{\mbox{elec}}$ does not produce an operator with a
discrete spectrum since this would just cause the mass polarisation
term to vanish and the effective electronic mass to become the rest
mass. As we have seen it leads to `\index{adiabatic}adiabatic divergence' \cite{FROL:99}.

It is thus not possible to reduce the molecular \index{Schr\"{o}dinger}Schr\"{o}dinger
equation to a system of coupled differential equations of classical
type for nuclei moving on \index{Potential Energy Surface}Potential Energy Surfaces as suggested by \index{Born}Born. An \textit{extra choice} of fixed nuclear positions must be made to give any discrete spectrum and normalizable $L^2$ eigenfunctions.  In our view this choice, that is, the introduction of the \index{clamped-nuclei} clamped-nuclei \index{Hamiltonian}Hamiltonian into the molecular theory as in \S \ref{BOtheory}, is the essence of what is commonly meant by the expression\footnote{In its original form ${\bf b}~=~{\bf b}_o$, the \index{equilibrium configuration}equilibrium configuration, on the right-hand side of (\ref{BOapprox}).}, the `\index{Born}Born-\index{Oppenheimer}Oppenheimer approximation' 
\begin{equation}
\mathsf{\hat{H}}^{\mbox{elec}}~=~  \int^{\oplus}_{X}\mathsf{\hat{K}}({\bf b},{\bf t}^{\rm e})_o~d{\bf b}~\rightarrow~\mathsf{\hat{K}}({\bf b},{\bf t}^{\rm e})
\label{BOapprox}
\end{equation}
If the molecular \index{Hamiltonian}Hamiltonian $\mathsf{H}$ were classical as in \cite{BH}, the removal of the nuclear \index{kinetic energy}kinetic energy terms would indeed leave a \index{Hamiltonian}Hamiltonian representing the electronic motion for \index{stationary nuclei}stationary nuclei, as claimed by \index{Born}Born and \index{Oppenheimer}Oppenheimer \cite{BO,BOBl}. As we have seen, quantization of $\mathsf{H}$ changes the situation drastically, so an implicit appeal to the classical limit for the nuclei is required. The argument is a subtle one, for subsequently, once the classical energy surface has emerged, the nuclei are treated as quantum particles for the determination of the vibration-rotation spectrum (though indistinguishability is rarely carried through); this can be seen from the complexity of the mathematical account given by Klein and co-workers \cite{KMSW:92}.

This \textit{qualitative} modification of the internal \index{Hamiltonian}Hamiltonian, the extra choice of fixed nuclear positions in the `electronic' \index{Hamiltonian}Hamiltonian, is \textit{ad hoc} in the same sense that \index{Bohr}Bohr's quantum theory of the atom is an \textit{ad hoc} modification of classical mechanics. An essential feature of the answer is put in by hand. We know that both modifications have been tremendously useful and our point is not that something else must be done in practical calculations on molecules. The point is how the successful description of molecules involving the \index{clamped-nuclei} clamped-nuclei modification at some stage can best be understood in terms of quantum mechanics. In the case of the \index{Bohr}Bohr atom the resolution of the inconsistency in mechanics applied to the microscopic realm was achieved quite quickly with the formulation of quantum mechanics; in the molecular case, no such resolution is at present known.

\section*{Appendix}
\addcontentsline{toc}{section}{Appendix}

In this appendix we give an heuristic account of the mathematical notion of the \index{direct integral}direct integral of \index{Hilbert space}Hilbert spaces, and then study a simple model problem to illustrate the general ideas discussed in the paper. 

Consider a \index{self-adjoint}self-adjoint operator $T$ that depends on a parameter $X$, so $T~=~T(X)$. The parameter $X~=~-\infty\leq~X~\leq+\infty$, covers the whole real line ${\cal{R}}$, and $T(X)$ is assumed to be defined for all $X$. $T(X)$ is an operator on a \index{Hilbert space}Hilbert space, which is denoted ${\cal{H}}(X)$; its eigenfunctions \{$\phi$\} defined by
\begin{displaymath}
T(X)~\phi(X)_k~=~\lambda_k(X) ~\phi(X)_k.
\end{displaymath}
form a complete orthogonal set for the space ${\cal{H}}(X)$. The scalar product is
\begin{equation}
\langle \phi(X)_{k}|\phi(X)_{j}\rangle_{X}~=~\int \phi(X,x)_{k}^{*}\phi(X,x)_{j}~dx~=~f(X)_{kj}~\equiv~f(X)~\delta_{kj}, f(X)~<~\infty
\label{scalp}
\end{equation}
 $T$ may have discrete eigenvalues below a \index{continuous spectrum}continuous spectrum that starts at $\Lambda$, that is
\begin{displaymath}
\sigma(X)~=~\sigma(T(X))~=~\left[\lambda_0(X),\lambda_1(X),\ldots,\lambda_k(X)\right)~\bigcup~\left[\Lambda(X),\infty\right).
\end{displaymath}

Now let's introduce a `big' \index{Hilbert space}Hilbert space ${\cal{H}}$ as a \textit{\index{direct integral}direct integral} over the \{${\cal{H}}(X)$\},
\begin{equation}
{\cal{H}}~=~\int^{\oplus}_{\cal{R}}~{\cal{H}}(X)~dX
\label{apdirint}
\end{equation}
and correspondingly the operator ${\cal{T}}$ acting on ${\cal{H}}$ defined by
\begin{displaymath}
{\cal{T}}~=~\int^{\oplus}_{\cal{R}}~T(X)~dX.
\end{displaymath}
The scalar product on the big space ${\cal{H}}$ is defined explicitly in terms of (\ref{scalp}) by
\begin{equation}
\langle \phi(X)_{k}|\phi(X)_{j}\rangle_{{\cal{H}}}~:=~\int_{{\cal{R}}}\langle \phi(X)_{k}|\phi(X)_{j}\rangle_{X}~dX~<~\infty
\label{bigscalp}
\end{equation}
In equation (\ref{scalp}) one can always chose the functions \{$\phi_k$\} to be orthonormalized independently of $X$,
\begin{displaymath}
f(X)~=~1
\end{displaymath}
However this choice is not consistent with (\ref{bigscalp}), which requires $f(X)$ to decrease sufficiently fast as $|X|\rightarrow \infty$ for the integral to exist. The mathematical motivation for this construction is this: the cartesian product of the spaces \{${\cal{H}}(X)$\},
\begin{displaymath}
{\cal{F}}~=~\prod_{X \in {\cal{R}}}~{\cal{H}}(X)
\end{displaymath}
is a field of \index{Hilbert space}Hilbert spaces over ${\cal{R}}$ which has a natural vector space structure. In modern geometric language, the \index{Hilbert space}Hilbert space ${\cal{H}}(X)$ is a fibre over a point $X$ in a fibre bundle; ${\cal{F}}$ is the vector space of sections of this bundle. The subspace of ${\cal{F}}$ consisting of \textit{square integrable} sections is the \index{direct integral}direct integral (\ref{apdirint}).
The \index{direct integral}direct integral is the generalization to the continuous infinite dimensional case of the notion of the direct sum of finite dimensional vector spaces.

Purely as a heuristic explanation suppose initially that the parameter $X$ has only two discrete values \{$X_1,X_2$\}. There are then two eigenvalue equations to consider, and two associated \index{Hilbert space}Hilbert spaces. In the direct sum space we have
\begin{displaymath}
{\cal{T}}_2~=~T(X_1)~\oplus~T(X_2),~~~{\cal{H}}~=~{\cal{H}}(X_1)~\oplus~{\cal{H}}(X_2)
\end{displaymath}
The eigenfunctions of ${\cal{T}}_2$ are then two-dimensional column vectors, and so
\begin{displaymath}
{\cal{T}}_2~\left[\begin{array}{c}\phi(X_1)_k\\\phi(X_2)_j\end{array}\right]~=~
\big(\lambda_k(X_1)~+~\lambda_j(X_2)\big)~\left[\begin{array}{c}\phi(X_1)_k\\\phi(X_2)_j\end{array}\right]
\end{displaymath}
The spectrum of ${\cal{T}}_2$ is the union of the spectra of $T(X_1)$ and $T(X_2)$. This discussion is trivially extended to $n$ points \{$X_k:k=1,\ldots,n$\}, with the spectrum of ${\cal{T}}_n$ given by the union of the $n$ operators $T(X_1),\ldots,T(X_n)$. The limit $n\rightarrow \infty$ is not trivial since it brings in important notions from topology and integration (measure theory) which we gloss over \cite{JD:81}. When these are taken into consideration however the result is that the spectrum $\sigma$ of ${\cal{T}}$ is \textit{purely continuous} since its \index{direct integral}direct integral representation implies that its spectrum is the union of the spectra of the infinite set of $T(X)$ operators,
\begin{equation}
\sigma~=~\sigma({\cal{T}})~=~\bigcup_X~\sigma({X})~\equiv \left[L_0,\infty\right)
\label{contspec}
\end{equation}
where $L_0$ is the \underline{minimum} value of $\lambda_0(X)$. The  eigenvalue equation for ${\cal{T}}$ is,
\begin{displaymath}
{\cal{T}}~\Phi_{\Lambda}~=~\Lambda~\Phi_{\Lambda},~~~~L_0~\leq~\Lambda~<\infty
\end{displaymath}
with $\Lambda$ a continuous index for the \{$\Phi$\}.  Even if $T(X)$ is \index{self-adjoint}self-adjoint, it doesn't follow that its \index{direct integral}direct integral ${\cal{T}}$ is \index{self-adjoint}self-adjoint; that depends on specifics and has to be investigated. So the \{$\Phi$\} cannot be assumed to be complete.

In the application of this mathematics to the \index{Born}Born-\index{Oppenheimer}Oppenheimer approximation, the role of $x$ is taken by the electronic coordinates ${\bf t}^{\rm e}$, and $X$ is to be identified with definite choices of the nuclear coordinates ${\bf b}$. If there are $M$ nuclei the parameters ${\bf b}$ are elements of ${\cal{R}}^{3(M-1)}$. The operator $T(X)$ is the \index{clamped-nuclei} clamped-nuclei \index{Hamiltonian}Hamiltonian $\mathsf{\hat{K}}({\bf b},\hat{\bf t}^{\rm e})_o~=~\mathsf{\hat{K}}_o$. With the conventional normalization of \index{clamped-nuclei} clamped-nuclei electronic eigenfunctions independent of the nuclear positions ${\bf b}$, the formal eigenvectors, (\ref{nonorm}), \cite{am60} of $\mathsf{\hat{H}}^{\mbox{elec}}$  do not belong to the \index{Hilbert space}Hilbert space ${\cal{H}}$; this simply reflects the loose use of the Dirac delta function for the \index{position operator}position operator eigenfunctions.

We now consider a concrete model consisting of coupled harmonic oscillators with two degrees of freedom; we try to mimic the steps taken in the usual \index{Born}Born-\index{Oppenheimer}Oppenheimer discussion.  Consider the following \index{Hamiltonian}Hamiltonian where $\kappa$ and $a$ are dimensionless constants\footnote{The variables are expressed in dimensionless form for simplicity. The quantum oscillator $\mathsf{\hat{h}}~=~\tfrac{1}{2}(\mathsf{\hat{p}}^2~+~\mathsf{\hat{q}}^2)$ has eigenvalues $n~+~\tfrac{1}{2}$.} 
\begin {equation}
\mathsf{\hat{H}}~=~\mathsf{\hat{p}}^2~+~\kappa^4\mathsf{\hat{P}}^2~+~\mathsf{\hat{x}}^2~+~\mathsf{\hat{X}}^2~+~a\mathsf{\hat{x}}\mathsf{\hat{X}}
\label{fullHam}
\end{equation}
The only non-zero commutators of the operators are
\begin{displaymath}
[\mathsf{\hat{x}},\mathsf{\hat{p}}]~=~i~~~[\mathsf{\hat{X}},\mathsf{\hat{P}}]~=~i
\end{displaymath}
Following the conventional discussion of electron-nuclear separation outlined in \S \ref{BOtheory},\S \ref{Brnadi}, define
\begin{eqnarray}
\mathsf{\hat{H}}_o~&=&~\mathsf{\hat{p}}^2~+~\mathsf{\hat{x}}^2~+~a\mathsf{\hat{x}}\mathsf{\hat{X}}~+~\mathsf{\hat{X}}^2\\
\mathsf{\hat{H}}_1~&=&~\mathsf{\hat{P}}^2
\label{ops}
\end{eqnarray}
so that
\begin{equation}
\mathsf{\hat{H}}~=~\mathsf{\hat{H}}_o~+~\kappa^4~\mathsf{\hat{H}}_1
\label{separate}
\end{equation}
with \index{Schr\"{o}dinger}Schr\"{o}dinger equation
\begin{equation}
\left(\mathsf{\hat{H}}~-~E\right)\Psi~=~0
\label{Scheq1}
\end{equation}
We note that
\begin{equation}
[\mathsf{\hat{H}}_o,\mathsf{\hat{H}}_1]~\neq~0
\label{noncomm}
\end{equation}
so the two parts cannot be simultaneously diagonalized. A principal axis transformation of the whole expression $\mathsf{\hat{H}}$ brings it to separable form, but we do not need to pursue explicitly the full solution here.

On the other hand
\begin{equation}
[\mathsf{\hat{H}}_o,\mathsf{\hat{X}}]~=~0
\label{comm}
\end{equation}
so these two operators may be simultaneously diagonalized, and consider $\mathsf{\hat{H}}_o$ at a definite eigenvalue of $\mathsf{\hat{X}}$, say $X$
\begin{equation}
\mathsf{\hat{K}}_o~=~\mathsf{\hat{p}}^2~+~\mathsf{\hat{x}}^2~+~a\mathsf{\hat{x}}{X}~+~{X}^2
\label{cnHam}
\end{equation}
This is the \index{Hamiltonian}Hamiltonian (in the variables $\mathsf{\hat{x}},\mathsf{\hat{p}}$) of a displaced oscillator in which $X$ is a (c-number) parameter, with \index{Schr\"{o}dinger}Schr\"{o}dinger equation in position representation
\begin{equation}
\mathsf{\hat{K}}_o~\varphi(x,X)_n~=~E_n(X)~\varphi(x,X)_n
\label{cnSch}
\end{equation}
$\mathsf{\hat{K}}_o$ is the analogue in this model of the `\index{clamped-nuclei} clamped-nuclei' electronic \index{Hamiltonian}Hamiltonian.

The solution is immediate; we make a unitary transformation by introducing a displaced coordinate involving $X$
\begin{eqnarray}
\mathsf{\hat{x}}'~&=&~\mathsf{\hat{x}}~-~\tfrac{1}{2}aX,~~~\mathsf{\hat{p}}'~=~\mathsf{\hat{p}}\\
\mathsf{\hat{U}}~&=&~e^{iaX\mathsf{\hat{p}}/2},~~~\mathsf{\hat{K}}_{o}^{'}~=~\mathsf{\hat{U}}^{-1}\mathsf{\hat{K}}_0 \mathsf{\hat{U}}
\label{unit}
\end{eqnarray}
so that the transformed $\mathsf{\hat{K}}_o$ in the new variables is
\begin{displaymath}
\mathsf{\hat{K}}'_{o}~=~\mathsf{\hat{p}}'{}^{2}~+~\mathsf{\hat{x}}'{}^{2}
~+~(1~-~\tfrac{a^2}{4})X^2
\end{displaymath}
which has oscillator eigenvalues and eigenfunctions
\begin{equation}
E_n(X)~=~E_n^0~+~(1~-~\tfrac{a^2}{4})X^2,~~~\varphi(x')_n
\label{eigen}
\end{equation}
where $x'~=~x~-~\tfrac{aX}{2}$, the \{$\varphi_n$\} are the usual harmonic oscillator eigenfunctions and $E_n^0$ is the energy of the free oscillator $2(n~+~\tfrac{1}{2})$. For fixed $n$ the spectrum $\sigma(X)$ is discrete and, as a function of the $X$ parameters, would be conventionally interpreted as a `potential energy curve'. As far as (\ref{separate}) is concerned, $\mathsf{\hat{K}}_o$, evaluated at some point $X_0$ \textit{cannot} be regarded as an `approximation' to $\mathsf{\hat{H}}_o$, since obviously
\begin{displaymath}
[\mathsf{\hat{K}}_{o}(X_0),\mathsf{\hat{H}}_1]~=~0
\end{displaymath}
so they can be simultaneously diagonalized, and $\mathsf{\hat{H}}_1$ has purely \index{continuous spectrum}continuous spectrum (free motion). So we have to consider $X$ in the full problem in its operator form, $\mathsf{\hat{X}}$.

We make the same unitary transformation of the $\mathsf{\hat{x}},\mathsf{\hat{p}}$ variables in $\mathsf{\hat{H}}_o$ as before, and it is still brought to diagonal form; however $\mathsf{\hat{H}}_1$ will be modified because $\mathsf{\hat{P}}$ is also translated by the operator $\mathsf{\hat{U}}$ in (\ref{unit}) that generates the coordinate displacement (cf equation (\ref{noncomm})); thus
\begin{displaymath}
\mathsf{\hat{U}}^{-1}~\mathsf{\hat{P}}~\mathsf{\hat{U}}~=~\mathsf{\hat{P}}~+~\tfrac{1}{2}a\mathsf{\hat{p}}
\end{displaymath}
so the transform of $\mathsf{\hat{H}}_1$ contains linear and quadratic terms in $\mathsf{\hat{p}}$.

Nevertheless (\ref{comm}) is still valid, and formally we may write $\mathsf{\hat{H}}_0$ as a \index{direct integral}direct integral
\begin{equation}
\mathsf{\hat{H}}_o~=~\int^{\oplus}_{X}\mathsf{\hat{K}}(\mathsf{\hat{x}},\mathsf{\hat{p}},X)_o~dX
\label{dirint1}
\end{equation}
The \index{Schr\"{o}dinger}Schr\"{o}dinger equation for $\mathsf{\hat{H}}_o$ in position representation is now an equation involving functions of $x$ and $X$
\begin{equation}
\mathsf{\hat{H}}_o~\Phi_{\epsilon}(x',X)~=~\epsilon~\Phi_{\epsilon}(x',X)
\label{hosch}
\end{equation}
Just as before (see (\ref{contspec})) the \index{direct integral}direct integral decomposition (\ref{dirint1}) implies that the spectrum is purely continuous, explicitly
\begin{equation}
\sigma(\mathsf{\hat{H}}_o)~=~\bigcup_{X} \sigma(X)~=~\big[1,\infty\big).
\label{spec}
\end{equation}
$\epsilon$ in (\ref{hosch}) takes all positive values $\geq~1$, where $1$ is the \textit{minimum} value of the oscillator eigenvalue $2(n~+~\tfrac{1}{2})$. The associated continuum eigenfunctions \{$\Phi$\} may formally be written as products of oscillator eigenfunctions (in $x'$), and delta functions (in $X$). They don't lie in \index{Hilbert space}Hilbert space of course and one needs the Gel'fand construction of a rigged space to make sense of the formal calculation \cite{LEB:90}. If one returns to the $x$ variable, the \{$\varphi_n$\} are functions of $x$ and $X$, since $x'~=~x'(X)$.

One can't do anything very useful with the \index{direct integral}direct integral expression (\ref{dirint1}) for $\mathsf{\hat{H}}_o$ apart from adding it onto the $\kappa^4$ term, which just returns us to the full problem. The full wavefunction in (\ref{Scheq1}) can be expanded as
\begin{eqnarray}
\Psi(x',X)~&=&~{\mathrlap{\sum}\int}_{n,X'} \varphi\big(x'(X')\big)~\delta(X~-~X')~c(X')_n~dX'\\
~&=&~\sum_n c(X)_n~\varphi\big(x'(X)\big)_n
\label{expans}
\end{eqnarray}
which obviously leads towards a variational approach \cite{am60}; such expansions rely on the completeness of the states employed. In this simple problem there is no difficulty, but as noted earlier, in realistic \index{Coulomb}Coulomb systems it is much less clear that a complete set of states is available. 

However that may be, let us rehearse again the argument due to \index{Born}Born summarized in \S \ref{Brnadi}. We substitute (\ref{expans}) in equation (\ref{Scheq1}), left multiply by $\varphi_{m}^{*}$ and integrate out the $x'$ variables to leave an equation for the coefficients \{$c(X)_n$\},
\begin{multline}
\int dx'~\varphi(\big(x'(X)\big)_{m}^{*}\bigg[\mathsf{\hat{H}}_o~+~\kappa^4~\mathsf{\hat{H}}_1\bigg]~\sum_{n}c(X)_n~\varphi(\big(x'(X)\big)_n~\\=~E~\int dx'~\varphi(\big(x'(X)\big)_{m}^{*}~\sum_{n}c(X)_n~\varphi(\big(x'(X)\big)_n
\label{Bornex}
\end{multline}
At this point in the conventional account, $\mathsf{\hat{H}}_o$ is replaced by $\mathsf{\hat{K}}_o$, equation (\ref{cnHam}), and then the action of $\mathsf{\hat{K}}_o$ on the functions \{$\varphi$\} in equation (\ref{Bornex}) can be evaluated using (\ref{cnSch}) in the well-known way,
\begin{displaymath}
\mathsf{\hat{H}}_o~\varphi\big(x'(X)\big)_n~\rightarrow~\mathsf{\hat{K}}_o~\varphi\big(x'(X)\big)_n~=~\epsilon(X)_n~\varphi\big(x'(X)\big)_n
\end{displaymath}
>From the foregoing discussion it is clear that the substitution of $\mathsf{\hat{H}}_o$ by $\mathsf{\hat{K}}_o$ makes a \textit{qualitative} change in the theory. This change does seem to be the `right' thing to do, but so far there is \textit{no} explanation as to \textit{why} this is so.

\end{document}